\documentclass[a4paper,fleqn]{cas-dc}

\usepackage{graphicx}
\usepackage{siunitx}
\usepackage{subfigure}
\usepackage{epstopdf}
\usepackage{url}
\usepackage[authoryear]{natbib}

\usepackage{bbding}
\newcommand\extrafootertext[1]{%
    \bgroup
    \renewcommand\thefootnote{\fnsymbol{footnote}}%
    \renewcommand\thempfootnote{\fnsymbol{mpfootnote}}%
    \footnotetext[0]{#1}%
    \egroup
}

\begin{document}


   \title[mode = title]{Solar flare observations with the Radio Neutrino Observatory Greenland (RNO-G)}

   \shorttitle{Solar flare observations with RNO-G}

   \shortauthors{RNO-G Collaboration}
    \author[1]{S. Agarwal}
    \author[2]{J.~A. Aguilar}
    \author[1]{S. Ali}
    \author[3]{P. Allison}
    \author[4]{M. Betts}
    \author[1]{D. Besson}
    \author[5,6]{A. Bishop}
    \author[7]{O. Botner}
    \author[8]{S. Bouma}
    \author[9,10]{S. Buitink}
    \author[8]{M. Cataldo}
    \author[11]{B.~A. Clark}
    \author[7]{A. Coleman}
    \author[1]{K. Couberly}
    \author[6]{S. de~Kockere}
    \author[6]{K.~D. de~Vries}
    \author[12]{C. Deaconu}
    \author[5]{M.~A. DuVernois}
    \author[7]{C. Glaser}
    \author[7]{T. Gl{\"u}senkamp}
    \author[7]{A. Hallgren}
    \author[13]{S. Hallmann}\cormark[1]\ead{authors@rno-g.org; steffen.hallmann@desy.de}
    \author[14]{J.~C. Hanson}
    \author[4]{B. Hendricks}
    \author[13,8]{J. Henrichs}
    \author[7]{N. Heyer}
    \author[1]{C. Hornhuber}
    \author[3]{K. Hughes}
    \author[13]{T. Karg}
    \author[5]{A. Karle}
    \author[5]{J.~L. Kelley}
    \author[2,6]{M. Korntheuer}
    \author[13,15]{M. Kowalski}
    \author[16]{I. Kravchenko}
    \author[4]{R. Krebs}
    \author[8]{R. Lahmann}
    \author[6]{U. Latif}
    \author[8]{P. Laub}
    \author[16]{C.-H. Liu}
    \author[17]{M.~J. Marsee}
    \author[13,8]{Z.~S. Meyers}
    \author[1]{M. Mikhailova}\cormark[1]\ead{masha.mikhailova@icecube.wisc.edu}
    \author[21]{C. Monstein}
    \author[10]{K. Mulrey}
    \author[4]{M. Muzio}
    \author[13,8]{A. Nelles}
    \author[18]{A. Novikov}
    \author[1]{A. Nozdrina}
    \author[12]{E. Oberla}
    \author[19]{B. Oeyen}
    \author[18]{N. Punsuebsay}
    \author[13,8]{L. Pyras}
    \author[7]{M. Ravn}
    \author[19]{D. Ryckbosch}
    \author[2]{F. Schl{\"u}ter}
    \author[6,20]{O. Scholten}
    \author[18]{D. Seckel}
    \author[1]{M.~F.~H. Seikh}
    \author[6]{J. Stoffels}
    \author[8]{K. Terveer}
    \author[2]{S. Toscano}
    \author[5]{D. Tosi}
    \author[6,9]{D.~J. Van~Den~Broeck}
    \author[6]{N. van~Eijndhoven}
    \author[12]{A.~G. Vieregg}
    \author[11]{A. Vijai}
    \author[12]{C. Welling}
    \author[17]{D.~R. Williams}
    \author[12]{P. Windischhofer}
    \author[4]{S. Wissel}
    \author[1]{R. Young}
    \author[8]{A. Zink}

\affiliation[1]{organization={University of Kansas, Dept.\ of Physics and Astronomy, Lawrence, KS 66045}, country={USA}}
\affiliation[2]{organization={Universit\'e Libre de Bruxelles, Science Faculty CP230, B-1050 Brussels}, country={Belgium}}
\affiliation[3]{organization={Dept.\ of Physics, Center for Cosmology and AstroParticle Physics, Ohio State University, Columbus, OH 43210}, country={USA}}
\affiliation[4]{organization={Dept.\ of Physics, Dept.\ of Astronomy \& Astrophysics, Center for Multimessenger Astrophysics, Institute of Gravitation and the Cosmos, Pennsylvania State University, University Park, PA 16802}, country={USA}}
\affiliation[5]{organization={Wisconsin IceCube Particle Astrophysics Center (WIPAC) and Dept.\ of Physics, University of Wisconsin-Madison, Madison, WI 53703}, country={ USA}}
\affiliation[6]{organization={Vrije Universiteit Brussel, Dienst ELEM, B-1050 Brussels}, country={Belgium}}
\affiliation[7]{organization={Uppsala University, Dept.\ of Physics and Astronomy, Uppsala, SE-752 37}, country={Sweden}}
\affiliation[8]{organization={Erlangen Centre for Astroparticle Physics (ECAP), Friedrich-Alexander-University Erlangen-N\"urnberg, 91058 Erlangen}, country={Germany}}
\affiliation[9]{organization={Vrije Universiteit Brussel, Astrophysical Institute, Pleinlaan 2, 1050 Brussels}, country={Belgium}}
\affiliation[10]{organization={Dept.\ of Astrophysics/IMAPP, Radboud University, PO Box 9010, 6500 GL}, country={The Netherlands}}
\affiliation[11]{organization={Department of Physics, University of Maryland, College Park, MD 20742}, country={USA}}
\affiliation[12]{organization={Dept.\ of Physics, Dept.\ of Astronomy \& Astrophysics, Enrico Fermi Inst., Kavli Inst.\ for Cosmological Physics, University of Chicago, Chicago, IL 60637}, country={USA}}
\affiliation[13]{organization={Deutsches Elektronen-Synchrotron DESY, Platanenallee 6, 15738 Zeuthen}, country={Germany}}
\affiliation[14]{organization={Whittier College, Whittier, CA 90602}, country={USA}}
\affiliation[15]{organization={Institut f\"ur Physik, Humboldt-Universit\"at zu Berlin, 12489 Berlin}, country={Germany}}
\affiliation[16]{organization={Dept.\ of Physics and Astronomy, Univ.\ of Nebraska-Lincoln, NE, 68588}, country={USA}}
\affiliation[17]{organization={Dept.\ of Physics and Astronomy, University of Alabama, Tuscaloosa, AL 35487}, country={USA}}
\affiliation[18]{organization={Dept.\ of Physics and Astronomy, University of Delaware, Newark, DE 19716}, country={USA}}
\affiliation[19]{organization={Ghent University, Dept.\ of Physics and Astronomy, B-9000 Gent}, country={Belgium}}
\affiliation[20]{organization={Kapteyn Institute, University of Groningen, PO Box 800, 9700 AV}, country={The Netherlands}}
\affiliation[21]{organization={Istituto ricerche solari Aldo e Cele Daccò (IRSOL), Faculty of Informatics, Università della Svizzera italiana (USI), CH-6605 Locarno}, country={Switzerland}}
\cortext[cor1]{Corresponding author}

  \begin{keywords}
      solar flares \sep
      radio telescopes \sep
      UHE neutrinos \sep
      calibration
  \end{keywords}

  \begin{abstract}
    The Radio Neutrino Observatory -- Greenland (RNO-G) seeks discovery of ultra-high energy neutrinos from the cosmos through their interactions in ice. The science program extends beyond particle astrophysics to include radioglaciology and, as we show herein, solar observations, as well.
    Currently seven of 35 planned radio-receiver stations (24 antennas/station) are operational. These stations are sensitive to impulsive radio signals with frequencies between 80 and 700 MHz and feature a neutrino trigger threshold for recording data close to the thermal floor.
    RNO-G can also trigger on elevated signals from the Sun, resulting in nanosecond resolution time-domain flare data; such temporal resolution is significantly shorter than from most dedicated solar observatories. In addition to possible RNO-G solar flare polarization measurements, the Sun also represents an extremely useful above-surface calibration source.

    Using RNO-G data recorded during the summers of 2022 and 2023, we find signal excesses during solar flares reported by the solar-observing Callisto network and also in coincidence with $\sim$2/3 of the brightest excesses recorded by the SWAVES satellite. 
    These observed flares are characterized by significant time-domain impulsivity. Using the known position of the Sun, the flare sample is used to calibrate the \hbox{RNO-G} absolute pointing on the radio signal arrival direction 
    to sub-degree resolution. We thus establish the Sun as a regularly observed astronomical calibration source to provide the accurate absolute pointing required for neutrino astronomy. 
 
  \end{abstract}
\date{Received 2024}
   \maketitle

\section{Introduction}
The Radio Neutrino Observatory in Greenland (\hbox{RNO-G}) experiment is currently under construction near Summit Station, with the goal of measuring astrophysical neutrinos with energies exceeding $10^{15}$ eV. In its final form, \hbox{RNO-G} will consist of an array of several hundreds of radio antennas embedded in the glacial ice of Greenland, sensitive to radio signals produced by an in-ice neutrino interaction. \hbox{RNO-G} is designed for nearly-continuous up-time, recording data whenever a radio signal can be identified above the (largely) thermal noise floor. The \hbox{RNO-G} frequency response covers the range 80 - 700 MHz, which also overlaps with solar radio-emission frequencies during flaring. Beyond its primarily particle astrophysics mission, \hbox{RNO-G} data provides insights into solar signals. Those signals must also be identified as possible contamination to searches for radio emissions from down-coming cosmic rays. In addition, signals from the Sun also offer considerable utility as a tool to calibrate the instrument. 

In this article we outline the instrumental capabilities of RNO-G, detail data taken during solar flares in 2022 and 2023, elaborate on time-domain characteristics of the signals, and show how solar signals are used by \hbox{RNO-G} for hardware verification. Fortunately, the construction of the \hbox{RNO-G} instrument has coincided with the most recent solar maximum, expected to peak in 2024.

\subsection{RNO-G}
The primary goal of \hbox{RNO-G} is the discovery of astrophysical neutrinos above PeV energies \citep{RNO-G:2020rmc}. These neutrinos are predicted to arise from interactions of ultra high-energy cosmic rays, particularly those particles having the highest known energy in the universe \citep{Berezinsky:1969erk,Stecker:1973sy}, either with material around their production sites \citep{Waxman:1998yy} or with the cosmic microwave background \citep{Engel:2001hd}. While firmly predicted, the neutrino flux level is as-yet unknown and a measurement will strongly constrain our understanding of the high energy universe \citep{Halzen:2002pg}. 

RNO-G will measure neutrinos by detection of radio emissions produced via the Askaryan mechanism \citep{Askaryan:1961pfb}. A neutrino interacting in ice produces a cascade of elementary particles; as the shower evolves, it acquires an overall negative charge owing to scattering of atomic electrons into the cascade, and depletion of shower positrons via annihilation. The (net negative) charge distribution produced in this cascade moves faster than the speed of light in the medium and thus gives rise to radio pulses of nanosecond-scale duration along a narrow cone-shaped emission profile \citep{Zas:1991jv}, which then propagate through the radio-transparent ice to the antennas of \hbox{RNO-G} \citep{Aguilar2022a}. 
The well-defined position of the Sun can be used to calibrate the absolute pointing for the arrival direction of any signal. Reconstructing the neutrino direction in sparsely instrumented radio-neutrino detectors is, however, complicated by the fact that only a small portion of the emission cone is seen by the antennas, so it is essential to also determine which part of the cone was observed. The latter can be constrained by the polarization, amplitude distribution, and frequency spectrum of the signal, which are the dominant sources of uncertainty for the angular resolution on the neutrino direction~\citep{ARIANNA:2020zrg,Plaisier:2023cxz}. For an overview of the in-ice radio technique and a review of other experimental efforts, see \citep{Barwick:2022vqt}.

\begin{figure}
    \centering
    \includegraphics[width=0.45\textwidth, trim={1cm 1cm 1cm 1cm},clip]{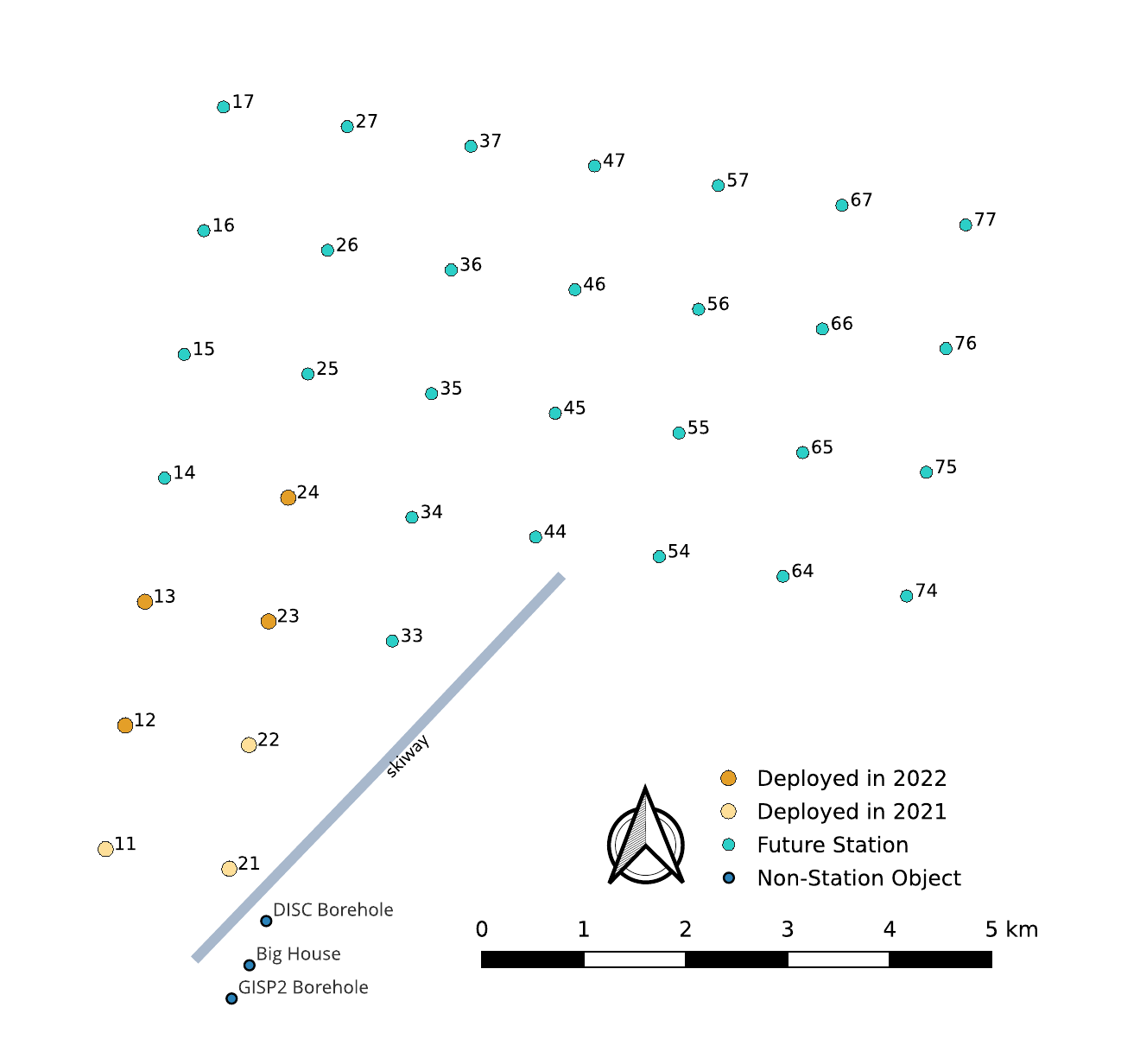}\newline
    \includegraphics[width=0.35\textwidth]{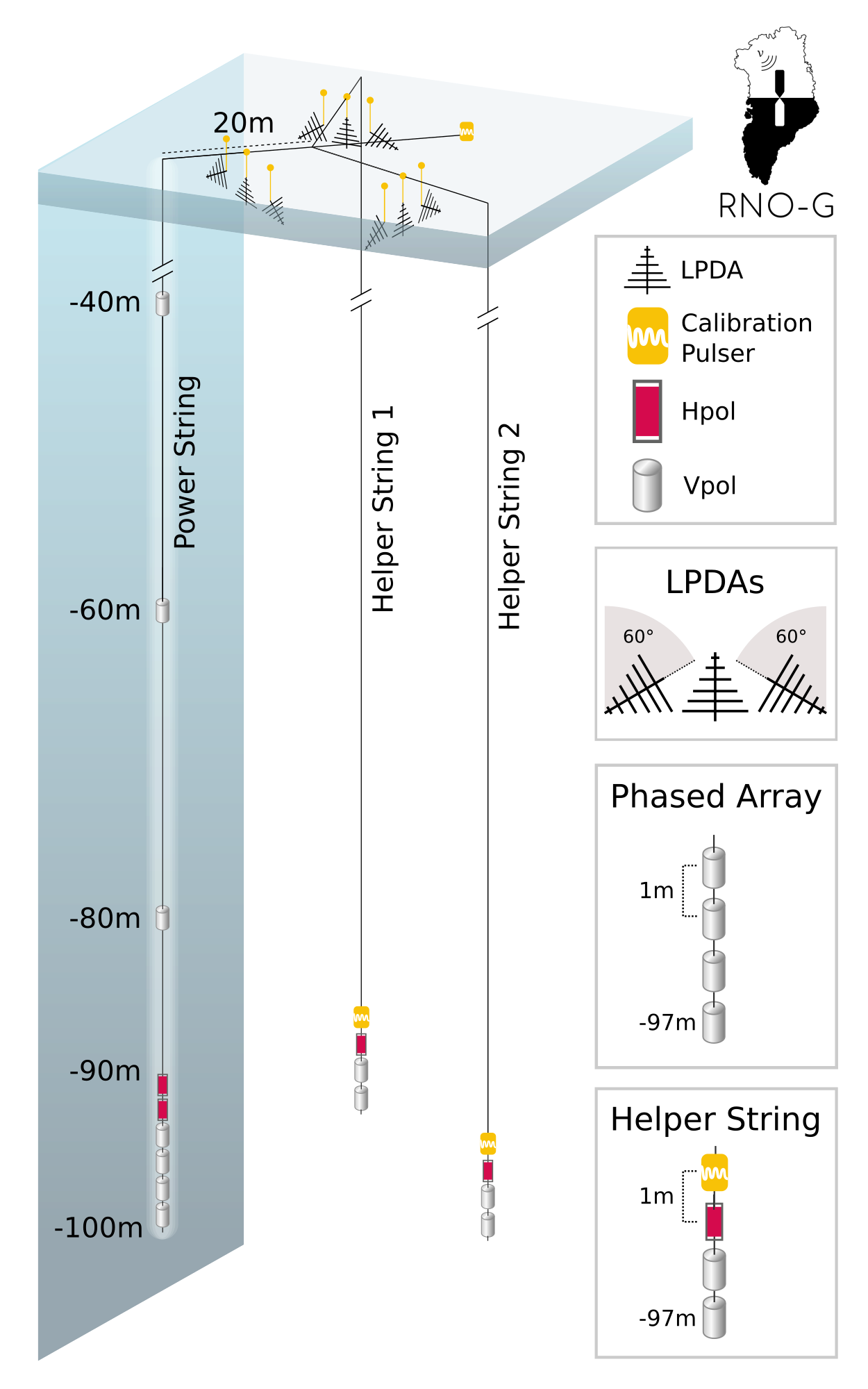}
    \caption{Top: Planned RNO-G station layout. The seven installed stations are highlighted in different colors. The main building of Summit Station is denoted by \emph{Big House}; also indicated is the landing strip for airplanes. Bottom: Schematic view of an \hbox{RNO-G} station. Each station has 24 antennas, either LPDA, Hpol or Vpol (see text for details). LPDAs are deployed in shallow trenches while Hpol and Vpol antennas are lowered into deep dry boreholes.}
    \label{fig:RNO-G-array}
\end{figure}

RNO-G is based on an array layout, whereby the planned 35 stations are installed on a square grid with interstitial spacing of 1.25 km (\autoref{fig:RNO-G-array}). Each station constitutes an independent and self-contained neutrino detector and consists of 24 antennas embedded in the ice. Each station combines log-periodic dipole antennas (LPDA) close to the surface with fat-dipoles (Vpol) and quad-slot antennas (Hpol) on instrumented strings down to 100\,m below the surface.
Construction began in 2021 with the commissioning of, and subsequent data collection for the first three stations. In 2022, four additional stations were added, for a total of seven stations active during the data collection relevant for this analysis. Construction will continue at least until 2026, adding new stations every year to reach 35 stations. Extensions beyond 35 stations are also possible.  

\begin{figure}
    \centering
    \includegraphics[width=\columnwidth]{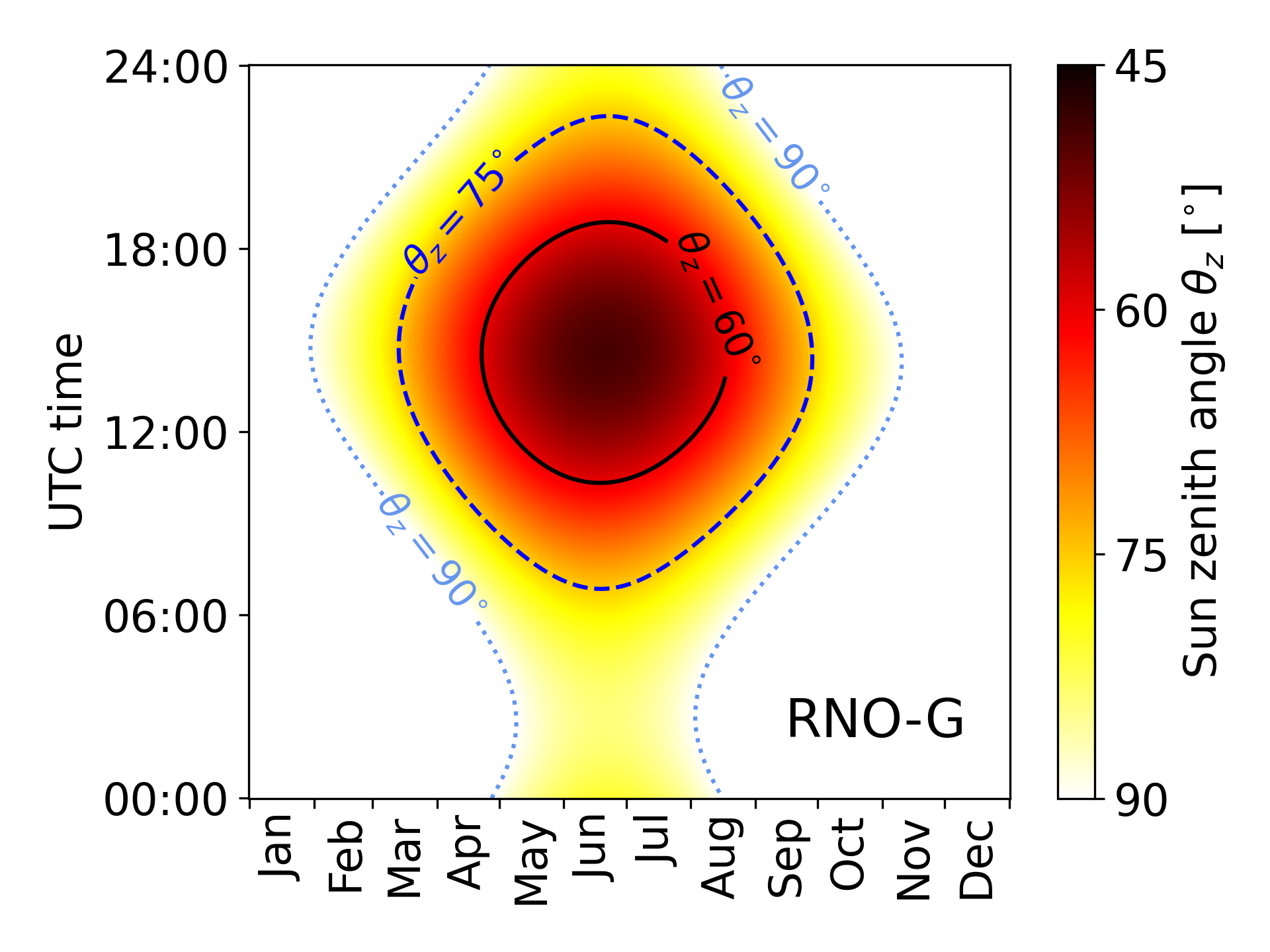} 
    \caption{Zenith position of the Sun observed from RNO-G. The Sun is above the horizon for the full day for 85 days of the year (May 10 to August 3). A minimum $\theta_z= 49^\circ$ is reached during solstice.}
    \label{fig:sun_zenith_rnog}        
\end{figure}

The stations are currently exclusively solar powered, such that the instruments turn off during the dark polar night (see \autoref{fig:sun_zenith_rnog}), although battery buffering ensures operation extending until October of each year. Since neutrino sensitivity scales directly with up-time, future wind-power operation should improve livetime with a targeted overall uptime of 80\%; solar observations, of course, are possible only when the Sun is above the horizon (cf. \autoref{fig:sun_zenith_rnog}).  

Following a trigger, recorded waveforms consist of 2048 samples written at a sampling rate of 3.2\,GHz, resulting in 640\,ns time windows for recorded events. (The sampling rate is planned to be lowered to 2.4\,GHz for future seasons to extend the time-period captured.) Given the sensitivity of antennas and this sampling rate, the stations have an effective bandwidth of 80\,MHz to 700\,MHz with a frequency resolution of better than 1\,MHz.  Station timing is determined via GPS clocks, meaning the station trigger times are, in principle, accurate to $\sim$10\,ns. 

There are several different types of triggers currently running on the \hbox{RNO-G} stations. The primary trigger, tuned to a target trigger rate of 1\,Hz, is designed to trigger on neutrino interactions close to the noise floor using signals from the four Vpol antennas at the bottom of the power string. An auxiliary trigger formed by combining signals from the near-surface LPDAs targets a trigger rate of approximately 0.1\,Hz. Thresholds on the separate trigger paths dynamically self-adjust by raising/lowering signal thresholds per channel to achieve the pre-set target trigger rate. Consequently, during periods of high external noise such as solar bursts, the trigger rate will drop after the initial onset, as the threshold is raised once per second until the target rate is attained.
In addition, a continuous and un-biased sample of \emph{forced trigger} events is written out at fixed time intervals every 10 seconds, which guarantees continual monitoring of the ambient background. 

In principle, the upward facing \hbox{LPDAs} are intrinsically more sensitive to solar emission than the in-ice antennas because they are upward pointing and thus have a substantially higher gain in the direction of the Sun, as compared to the antennas embedded deep in the ice. However, since the main science goal of \hbox{RNO-G} is neutrino detection, the LPDA-based trigger is only auxiliary and therefore not the main driver of the data-rate.

\subsection{Solar Observations with RNO-G}
\label{sec:sun_rnog}
Solar flares with accompanying radio-frequency emissions may arise via large prominences, or during periods of coronal mass ejections (CME). While the total energy output of the Sun is constant to within 0.1\% \citep{2016JSWSC...6A..30K} (dominated by infrared and optical wavelengths), the instantaneous power released in the radio and X-ray bands in a solar flare may exceed the background level by several orders of magnitude.

Observationally, solar flares are often classified via their X-ray emission, which are sorted according to their peak strength A, B, C, M, and X, with A being the smallest and X being the largest. Although related, there is no unambiguously direct correlation between X-ray and radio flares \citep{2011SSRv..159...19F}. Solar radio bursts are divided into five general classes I--V, based on their spectral and temporal characteristics; within these classes individual impulsive events exhibit considerable variation, e.g.~\cite{Dulk1985, Bastian:1998xq,1950AuSRA...3..387W}. 

Type I bursts are related to active hot-spots on the solar surface. They are short (second-long) quasi-monochromatic spikes with huge variation in intensity over the 80--200\,MHz frequency range. They usually cluster as `noise storms' which can last from a few hours up to a few days and sometimes feature a continuous underlying broadband emission. Although single signals show no eminent drift in frequency, the temporally spiking emissions tend to evolve towards lower frequencies during a storm.

Type II bursts are associated with electrons accelerated in the shock fronts at the leading edge of CMEs. They last a few minutes during which the narrow-band emission of the fundamental (harmonic) mode slowly drifts from $\sim$150 (300)\,MHz to lower frequencies. Typical drift rates are $\sim$0.05 MHz/s, consistent with a shock wave propagating outwards from the solar surface at $\sim$1000 km/s \citep{kumari23}.

Type III bursts are emitted by relativistic ($\sim$ 0.1--0.3\,$c$) electron beams travelling along open magnetic field lines during CME events. Compared to Type II, they are shorter (durations of a few seconds) and show a faster drift from up to $\sim$1\,GHz towards lower frequencies (up to $\sim$100 MHz/s). Type III bursts may occur in isolation or in clusters.
Rarely, Type III bursts do not continue to drift to lower frequencies towards the end of the flaring period, but show an inverted `J' or `U' shape in their dynamic spectra. For these `J' and `U' subtypes, electrons do not escape along open magnetic field lines but instead propagate downwards along a closed loop within the solar corona~\citep{1958Natur.181...36M}.

The most relevant types of bursts for \hbox{RNO-G} are the most abundant type III and II based on their frequency content and duration. While type II bursts are long enough to appear in even the unbiased event sample of forced triggers, type III bursts need to surpass the trigger threshold to be identified in the data stream.

The two remaining types IV and V occur only rarely and never in isolation. They are not easily identified in \hbox{RNO-G} due to their broad-band structure and because of their long duration or low frequencies, respectively.
Type IV flares comprise a smooth continuum of broad-band bursts, beginning 10--20 minutes after the flare maximum of some major flare events, with hour-scale duration.
Type V bursts are short-lived continuum noise at the lower frequency tail below 200\,MHz that, in rare cases, follow a Type III burst.

The sample of 2535 reported Callisto bursts (see below) in the field of view for \hbox{RNO-G} in the 2022 and 2023 data is dominated by type III (89\%) flares, followed by type II (4\%), V (2\%), IV (1\%), and other miscellaneous RF noise fluctuations (4\%). None of the few reported type I flares occurred in the \hbox{RNO-G} field of view during active data taking. 

For RNO-G, the salient features of solar flare emissions are as follows. Radiation is emitted by relativistic electron beams (type III bursts) or travelling shock waves (type II bursts). The emission is at the electron plasma frequency (Langmuir wave frequency)
\begin{equation}
        f_{pe} = \frac{1}{2\pi}\sqrt{\frac{e^2 n_e}{ m_e \epsilon_0}} \approx 0.009 \sqrt{\frac{n_e}{\mathrm{cm}^3}}\,\mathrm{MHz}
        \label{eq:plasma_frequency}
\end{equation}

or its harmonics. Several models exist that describe the electron density as a function of height in the solar corona (extending into the solar wind and interplanetary space), e.g.\ \cite{Mann1999}. Guided by such models, it is possible to (roughly) translate an observed frequency to a height above the solar surface. Using \autoref{eq:plasma_frequency} and the electron density profile, observed frequency drift rates $\mathrm{d}f$/$\mathrm{d}t$ can be converted to radial propagation velocities of the burst as well as elevation above the photosphere.

Generally, the emitted frequencies are lower further away from the Sun, and drift rates decrease faster at high frequencies due to the decreasing scale height of the plasma density as a function of solar radius. Given the lower frequency cutoff of the \hbox{RNO-G} amplifiers at $\sim\SI{80}{MHz}$, radio signals beyond $\sim1.4$ solar radii are not expected to be reconstructable in RNO-G.

The angular diameter of the solar photosphere is $0.54^\circ$ on the sky, although the solar spot, as viewed by \hbox{RNO-G} in-ice antennas, compresses in elevation due to refraction at the ice-air boundary. Radio emission in the frequency range relevant for \hbox{RNO-G} is contained within a $\sim1^\circ$ diameter. Hence, for RNO-G, radio pulses from the 149 million kilometers distant Sun arrive as a planar wave front from an infinitely far point-emitter, providing a unique calibration source to validate the pointing accuracy of the instrument.

Although plasma emission is the accepted mechanism behind type II and III solar flares and should imply strong circular polarization of the signal, type II and III typically show at most mild levels of circular polarization \citep{1958AuJPh..11..201K, wentzel1984}.

\subsection{Extant Solar Observatories}
Dedicated instruments continuously monitoring the Sun at radio frequencies include, among many others, the ground-based distributed Callisto network \citep{2009EM&P..104..277B} and satellite-based instruments like the WAVES instrument aboard the WIND satellite close to L1 \citep{1995SSRv...71..207L}, and the SWAVES instrument aboard the STEREO-A satellite \citep{2005AdSpR..36.1483K}.

To date, the Callisto network consists of more than 220 individual instruments distributed around the globe, of which on average more than 80 provide public (open source) radio spectrograms from continuous solar monitoring on a daily basis \citep{callisto_data}. Their data contain sweeps of 200 frequencies taken in 0.25\,s intervals in the native frequency range covering 45$\to$870 MHz, while some dedicated instruments cover 10$\to$80 MHz and others 1045$\to$1600 MHz using heterodyne converters. We find that the best overlap with \hbox{RNO-G} in global position, data availability, and frequency range is provided by the \hbox{HUMAIN} observatory in Belgium \citep{HUMAIN}, which is one of the instruments in the Callisto network. 

The SWAVES instrument is part of the STEREO mission \citep{2008SSRv..136..487B}, which consisted of two satellites revolving around the Sun, either ahead of (\mbox{STEREO-A}) or behind (STEREO-B) Earth's orbit. Although the \hbox{STEREO-B} mission concluded in 2014, STEREO-A is still operational and provides publicly accessible averaged spectrograms with minute binning that cover frequencies up to 16\,MHz. STEREO-A's orbital velocity is faster than Earth's; it lagged behind Earth by $50^\circ$ when \hbox{RNO-G} started operation in June 2021, overtook Earth in August 2023 and will lead by $50^\circ$ by 2026. A large fraction of the solar disk is therefore simultaneously observed by both SWAVES and also from terrestrial observatories.

There are several other dedicated solar observatories, such as the instruments flying on-board the Parker Solar Probe, e.g.~\cite{2016SSRv..204...49B}. Their data, however, was less readily available as a reference for \hbox{RNO-G}. Dedicated solar observatories, both existing as well as proposed e.g.~\cite{2023BAAS...55c.123G}, naturally have more targeted abilities to monitor the Sun. By contrast, although solar science is ancillary for instruments such as LOFAR~\citep{vanHaarlem:2013dsa}, the capabilities when dedicated observing time is used to point at the Sun can provide unprecedented detail e.g.~\cite{2019NatAs...3..452M}. These instruments are however not specialised in delivering continuous (almost real-time) data for identifying radio flares in the Earth's field of view.

\begin{figure*}
    \centering
    \includegraphics[width=\linewidth]{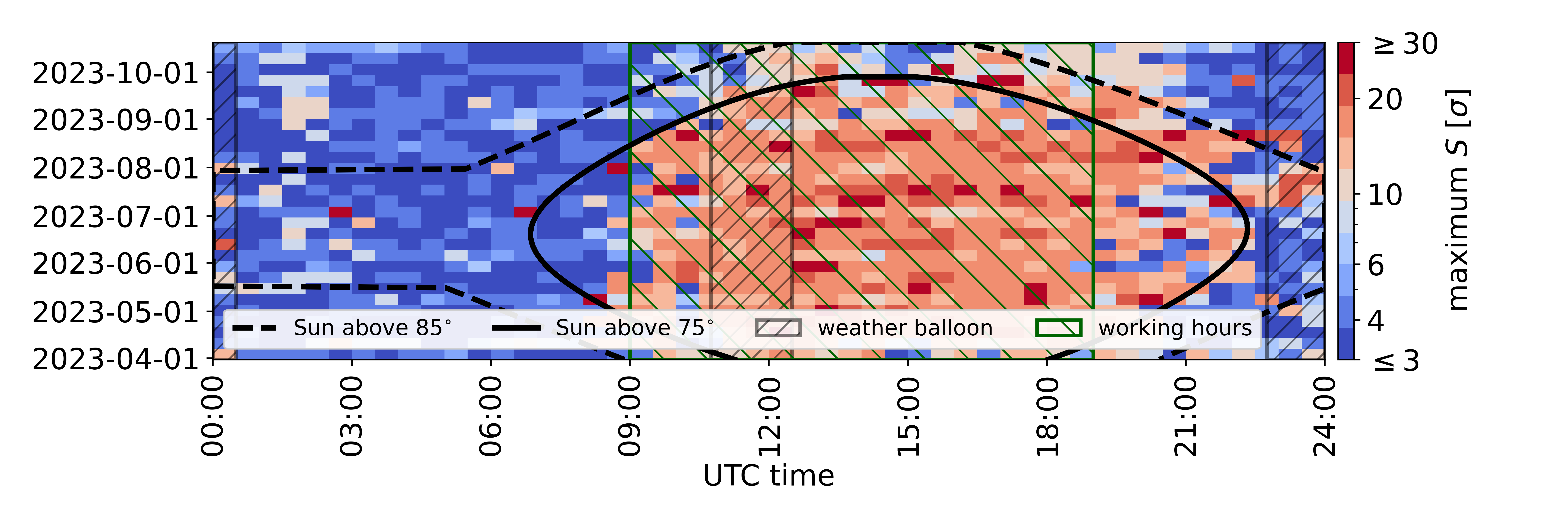}
    \caption{Deep trigger rate excesses (maximum recorded per week and daily half hour) observed in the \hbox{RNO-G} array as a function of time-of-day and time-of-year, illustrating the correlation with solar elevation. We note no obvious increase in trigger rates during balloon flights (grey hatched areas), but a clear enhancement of excesses during sunlight hours. However, most trigger rate enhancements are due to sources other than the Sun, given that human activity is also maximal at such times. This is particularly visible in the bias towards local evening hours. Normal working hours at Summit Station last from $\sim$7\,am to $\sim$5\,pm local time (green hatched areas).}
    \label{fig:trigger_rate_excess_sigmas}
\end{figure*}

\section{Identifying solar flares in RNO-G}
The \hbox{RNO-G} instrument, although not originally designed for solar observations, nevertheless has several unique capabilities that can inform our understanding of flares. Primary among these is the high (3.2 GSa/s) sampling rate and broadband frequency response (80 - 700 MHz), favoring data collection in the time domain. By contrast, typical solar observatories integrate power over a range of pre-defined frequencies in time increments of order 1-10 ms. Other radio particle arrays have previously reported solar flares, e.g.~\cite{Huber:2013wxf,ARIANNA:2015uwx,Clark:2019ady}, but either with much smaller bandwidth or in periods of low solar activity. \hbox{RNO-G} can therefore uniquely probe impulsive radio emissions on sub-nanosecond time scales over a very large frequency range. Since \hbox{RNO-G} strives for a continuous up-time to maximize neutrino livetime, the reliance on solar power provision ensures that \hbox{RNO-G} will be sensitive to flares whenever the Sun is above the horizon. 

We now discuss our identification of solar flares and analysis of those events.

\subsection{Elevated trigger rates}
Rather than continuously writing data to disk, \hbox{RNO-G} does so only when a trigger is issued. To prevent high-rate noise backgrounds from clogging the \hbox{RNO-G} data stream, the trigger thresholds are self-adjusting to keep a constant trigger rate. Several sources of transient noise phenomena that are capable of triggering the stations include:
\begin{itemize}
    \item Noise sources originating from the station itself, which constitute part of the `ambient' background and are not expected to lead to elevated trigger rates in the stations. Such noise sources include self-induced noise from the stations' power and communication systems, which was present in early data but was suppressed with subsequent hardware and electronics improvements.
    \item Tribo-electric discharge during periods of high wind periods are known to cause elevated trigger rates \citep{Aguilar:2023tba}.
    \item The radiosonde from twice-daily weather balloon flights telemeters data at 403 MHz. Depending on the flight trajectory, the weather balloon signals occasionally can elevate the trigger rates of individual stations.
    \item Flights to/from Summit Station and commercial airplanes flying overhead emit unintentional and intentional (such as radio altimeters to determine their height above ground) radio signals. Depending on height and proximity, these can trigger multiple stations.
    \item Local Radio-Frequency Interference (RFI) generated by human activity slightly raises the overall background rate for the 1--2 stations closest to the main building of Summit Station (or whenever field work is conducted at other locations), but its impact will diminish with increasing distance.
\end{itemize}

The lowest-level indicator of solar flaring activity is, therefore, a coherent increase in the trigger rates (as well as the recorded RMS voltages) across all stations.
Each \hbox{RNO-G} station has two main trigger conditions, corresponding to trigger formation in either the deep Vpol antennas (2/4 majority coincidence for the four deepest dipoles on the power string, see \autoref{fig:RNO-G-array}) or the surface \hbox{LPDA} antennas (2/3 majority logic for the three upward-pointing LPDAs).
To identify a sharp increase in the overall station trigger rates characteristic of solar flares, we use the excess metric given by \cite{Li:1983fv},
\begin{multline}
    S = \sqrt{2}\left[ N_\mathrm{on} \cdot \ln{\left(\frac{(1+\alpha)\,N_\mathrm{on}}{\alpha\,(N_\mathrm{on} + N_\mathrm{off})}\right)} \right.+ \\ \left. N_\mathrm{off} \cdot \ln{\left(\frac{(1+\alpha)\,N_\mathrm{off}}{N_\mathrm{on} + N_\mathrm{off}}\right)}\right]^{1/2}
    \label{eq:LiMa}
\end{multline}
to define a summed trigger rate excess in a sliding time window, where $N_\mathrm{on}$ is the summed number of events in all stations over a three second time interval, $N_\mathrm{off}$ is the local background rate of all stations over a 5 minute window centered on the on-time, and 
$\alpha=0.01$ the ratio of signal and background integration time. In the ideal case of zero signal contribution in the off-region, the value of $S$ would thus correspond to the excess significance, in units of $\sigma$, over the ambient background \citep{Li:1983fv}.
Under-fluctuations in the local background rate, e.g. due to stalled data-taking or the dynamic threshold adjustments, are masked by the daily average background rate to avoid artificial increases of the excess metric $S$.
The excess metric was formed separately for the surface \hbox{LPDA} antenna trigger vs. the in-ice Vpol trigger, and showed higher variability due to anthropogenic backgrounds for the surface component. The surface trigger rate was therefore not used for selecting our solar flare sample.  The maximum values obtained for the excess metric $S$ as a function of time-of-day and day of the year for the in-ice Vpol antenna triggers are shown for the 2023 season in \autoref{fig:trigger_rate_excess_sigmas}. It is evident that event trigger clusters are more prominent during the daytime, presumably linked to human activity; this, of course, coincides with times when the Sun's sky location reaches its maximum elevation above the horizon. Therefore, elevated trigger rates, although expected, are not a sufficient criterion to identify solar flares in RNO-G. More sophisticated search strategies are needed, in particular exploiting temporal correlations with data from dedicated solar observatories. 

\begin{figure*}
    \includegraphics[width=0.49\textwidth]{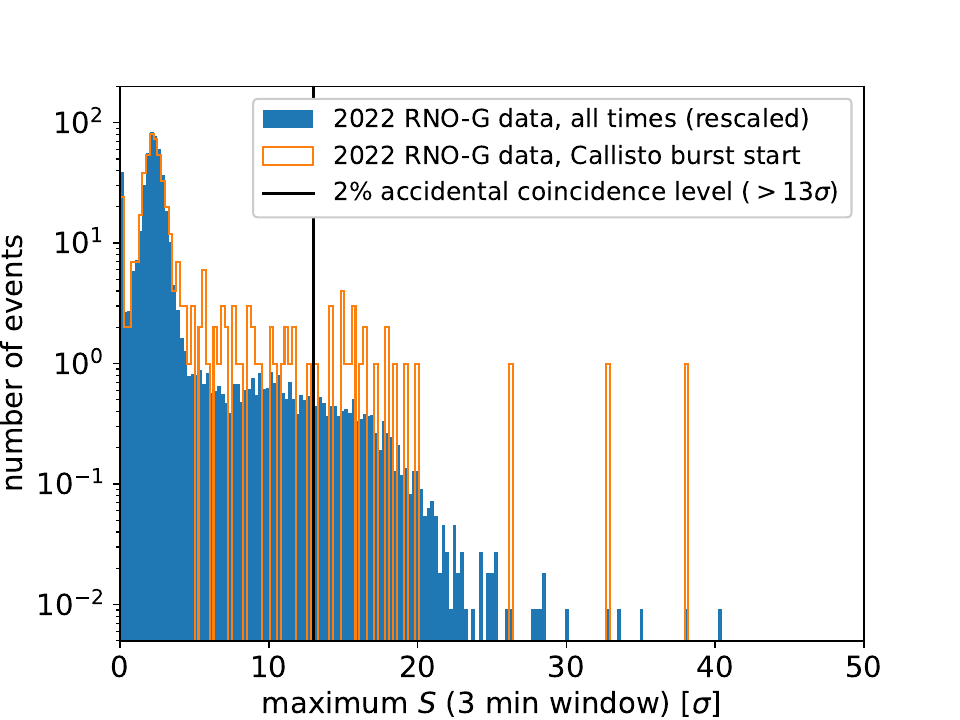}
    \hfill
    \includegraphics[width=0.49\textwidth]{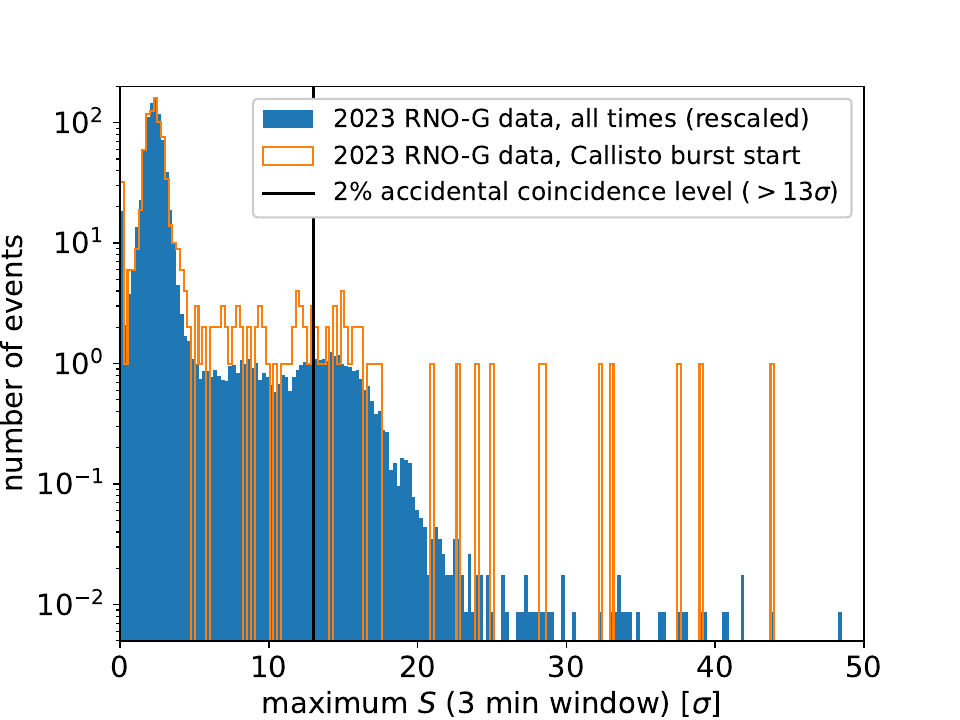}
    \caption{Deep trigger rate excess for 2022 (left) and 2023 (right) in all \hbox{RNO-G} stations during three minute periods around solar flares identified by the Callisto network. The cut indicates a 2 percent chance of a random excess generated by different backgrounds.}
    \label{fig:flare_stats}
\end{figure*}

\subsection{Correlations with dedicated solar observatories}
We use the trigger rate excess metric $S$ from  \autoref{eq:LiMa} to search for solar flare events in the \hbox{RNO-G} data and estimate the efficiency with which solar flare events are recorded by the \hbox{RNO-G} stations.
For this purpose, we use publicly available data from Callisto \citep{2009EM&P..104..277B} and SWAVES \citep{2008SSRv..136..487B}.

As shown in \autoref{fig:flare_stats}, we investigated all trigger rate excesses $S$ in \hbox{RNO-G} over 3 minute time-windows. Without considering any correlation with solar events, the maximum excesses for the full 2022 and 2023 data sets (respectively) follow the distribution shown in blue. Callisto publishes a list of identified burst times and observatories that observed an excess in their spectrograms. Burst times are reported with minute-precision, which motivated the 3 minute duration of the time window for our analysis. Selecting those periods where the calculated excess exceeds $S>13\sigma$ for all Callisto bursts identified by \hbox{HUMAIN} (orange) retains only 2\% of this distribution and allows us to identify a solar burst-enhanced sample. It is found that the \hbox{HUMAIN} referenced sample has a fractionally larger number of high trigger rate excursions.
The predefined 2\% accidental coincidence level cut is surpassed by 26 \hbox{HUMAIN} flares in 2022 and 49 flares in 2023.
Neglecting the few per-cent contamination in the classification of events, we infer that \hbox{RNO-G} recorded a total of 75 flares that caused a statistically significant number of triggers in the summers of 2022 and 2023.

To strengthen this conclusion, an additional correlation with SWAVES is performed. Although recording data only at a maximum frequency of 16\,MHz, and therefore below the nominal \hbox{RNO-G} turn-on frequency response, the instrument has the advantage that it continuously monitors the Sun with constant sensitivity and is free of transient local RFI backgrounds.
Hence, a sample of the largest SWAVES excesses (>15 dB) is selected and cross-matched with \hbox{RNO-G} to get an estimate for the detection efficiency of bright solar flares. In the complete 2022 and 2023 data, the SWAVES minute-binned data showed 119 excesses beyond 15\,dB at 16\,MHz during which \hbox{RNO-G} was actively recording data. Out of these, 53 flares showed an excess in the \hbox{RNO-G} trigger rates coincident with the flare time.

\begin{figure*}
    \centering
    \includegraphics[width=.49\textwidth]{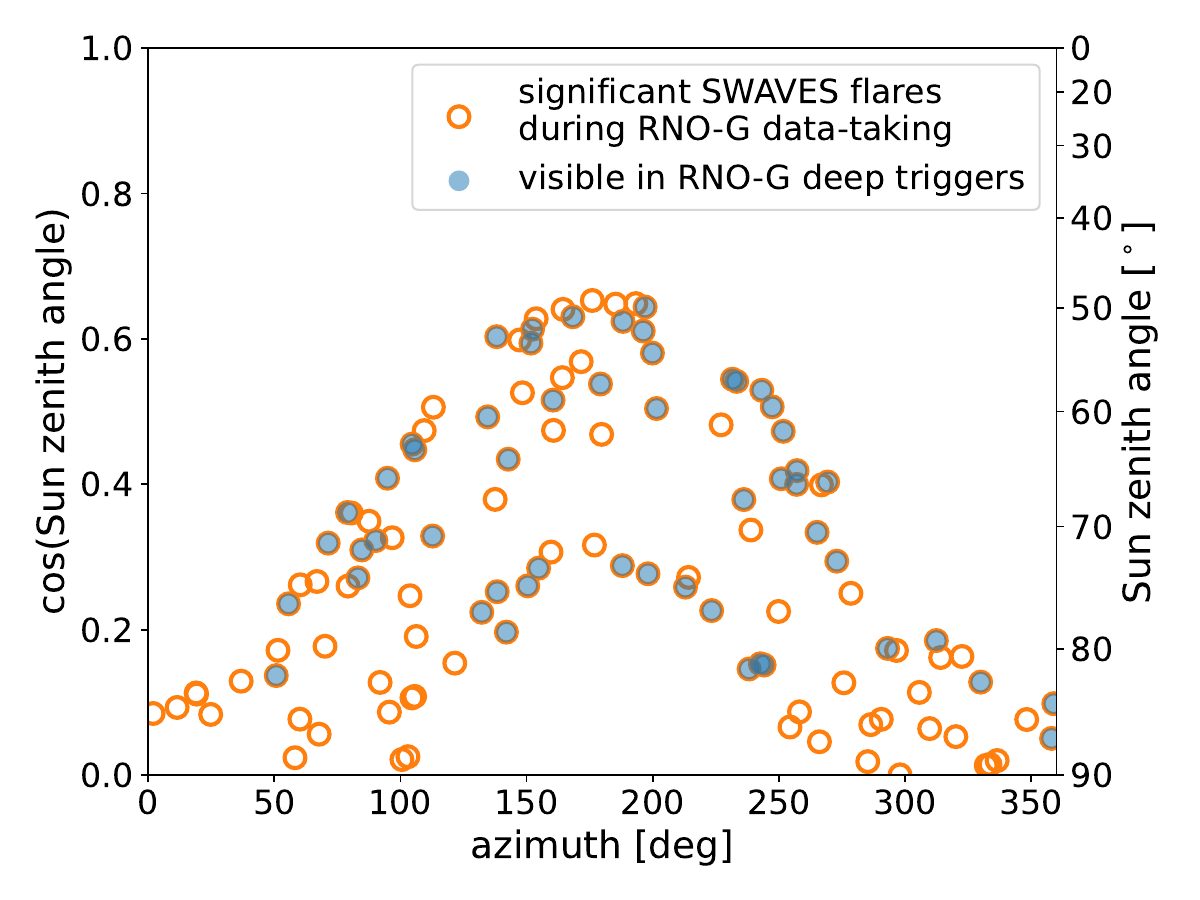}
    \hfill
    \includegraphics[width=.49\textwidth]{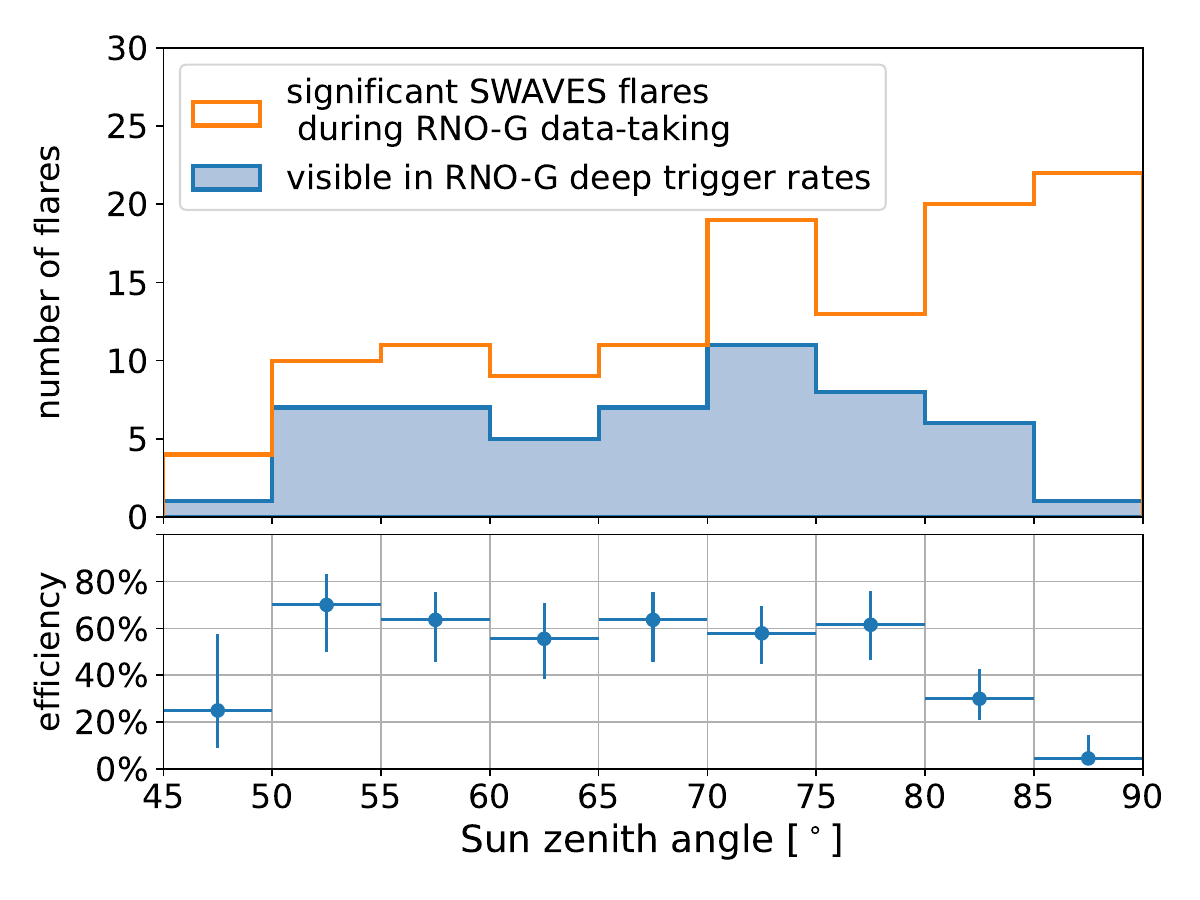}
    \caption{Left: Flares detected in SWAVES (open circles) and \hbox{RNO-G} (filled circles) as a function of local zenith and azimuth at the \hbox{RNO-G} site. 
    Right: Number of flares as function of local zenith. The orange lines show all flares that \hbox{RNO-G} could, in principle, have observed, while the blue curve shows the flares for which \hbox{RNO-G} observed an excess in the trigger rates. The \hbox{RNO-G} solar flare triggering efficiency in different regions of local zenith is shown below.}
    \label{fig:flare_stats_2}
\end{figure*}

As shown in \autoref{fig:flare_stats_2}, \hbox{RNO-G} detects a rise in trigger rates in roughly 60\% of solar flares that occur when the Sun is higher than $10^{\circ}$ above the horizon. For lower solar elevation angles, the efficiency drops, which can be explained by the fact that the Fresnel coefficients for transmission into the ice decrease significantly at grazing incidence angles. The Sun reaches zenith angles below 50$^\circ$ only for a short period close to summer solstice. The lower efficiency in the corresponding bin in \autoref{fig:flare_stats_2} is not significant because of the small statistics of flares in our sample. 

We did not observe an increase in efficiency when choosing even larger SWAVES excesses (beyond 15 dB).  We note that SWAVES flares without any counterpart at the higher RNO-G frequencies can have an impact on our total $\sim$60\% efficiency. However, for flares that originate close to the Sun and propagate outwards, radio emission will be present to some extent also at the higher RNO-G frequencies. Due to the different bands and locations of SWAVES and RNO-G, a direct comparison of flare strength and \hbox{RNO-G} trigger likelihood on an individual event basis is challenging. The key finding in the correlation with SWAVES excesses is therefore the relative drop in flare detection efficiency towards the horizon.

\subsection{Event signatures}
Solar flares with strong radio-frequency emissions have been historically classified and distinguished by their dynamic spectral characteristics and duration. 

\begin{figure*}
    \centering
    \includegraphics[width=0.85\linewidth]{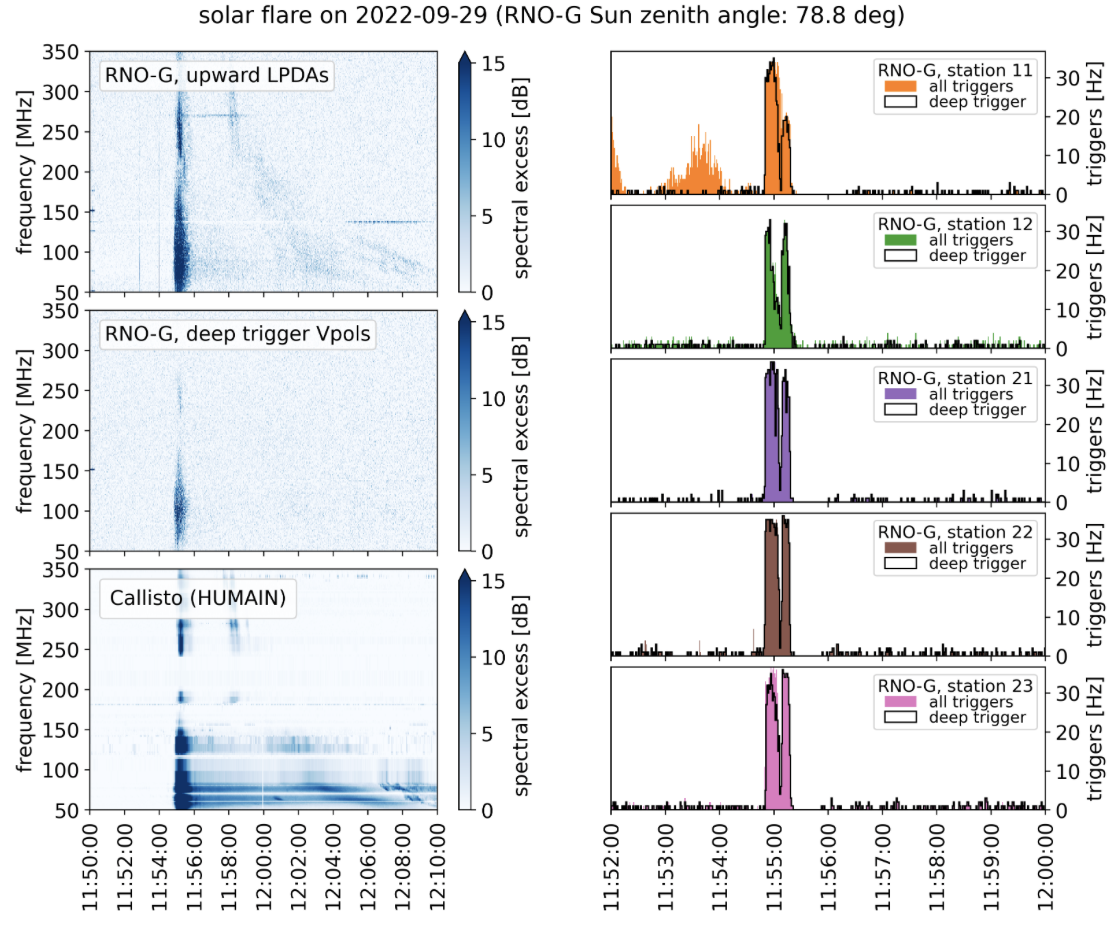}
    \caption{\textit{Left:} Frequency spectra from all active \hbox{RNO-G} stations observed in the surface \hbox{LPDA} channels (top) and in the $\sim$100\,m deep Vpol antennas used for triggering (bottom), as well as from the Callisto observatory \hbox{HUMAIN} (Belgium) during the 2022-09-29 solar flare. The spectral excess shown represents an excess over the median per frequency value. \textit{Right:} Deep and shallow trigger rates in all stations taking data during the flare. In all cases, the trigger rates quickly saturate at the flare onset; rates relax slightly as the flare temporarily weakens and the trigger servo raises the station trigger thresholds, then saturate again as the flare re-intensifies.}
    \label{fig:solar_flare_ICRC}
\end{figure*}

The \hbox{RNO-G} data contain a complete temporal record of each triggered event, which is thus more information than typical dynamic spectra, for which the phase information is discarded. In principle, \hbox{RNO-G} data could be used to independently search for solar flares by calculating continuous dynamic spectra, as is standard for other solar observatories. However, due to noise sources, as discussed above, this is not trivial and the identification of the flares considered in this article was facilitated (albeit not required) by association of \hbox{RNO-G} data with data from dedicated solar observatories. 
However, a dedicated \hbox{RNO-G} solar reconstruction pipeline would be necessary as dynamic spectra over time-scales of minutes are not characteristics of neutrino searches, where nanosecond-scale excursions are relevant. 

We do note that solar flares have been found in algorithmic studies for neutrino searches. Using for instance the machine learning method of anomaly detection \citep{RNO-G:2023psm}, it was found that waveforms recorded during solar bursts are considered anomalous compared to thermal noise dominated waveforms. A dedicated model was trained that identified a number of additional solar bursts (when comparing with data from other observatories), but the method also produced a large number of false positives (i.e.\ anomalies not related to the Sun due to the small set of labelled training data), so that it was not considered a conclusive identification strategy at this point for stand-alone solar analyses. Nevertheless, such an approach may hold promise after training on more detected flares in the future.   

The brightest flare observed in \hbox{RNO-G} thus far, which is easily identified by an elevated trigger rate, as well as in the anomaly search (or even by eye), is shown in \autoref{fig:solar_flare_ICRC}. This figure shows both the dynamic spectra of the LPDAs (more sensitive to signals coming from above) and the $\sim$100\,m deep Vpol antennas used for triggering. The time-dependence of the enhanced trigger rate is shown, which coincides precisely with the expectation from HUMAIN. The LPDAs register both the strong initial Type III burst, as well as the subsequent Type II ring-down of the frequency spectrum. The deep antennas, however, only register the initial burst, since the signal both has to traverse \SI{100}{m} of ice and also reaches the antennas in an unfavorable geometry.  

We provide a gallery for the subset of identified flares which we use for reconstruction in \autoref{sec:reconstruction} at \href{https://rnog-data.zeuthen.desy.de/solarflares}{https://rnog-data.zeuthen.desy.de/solarflares}. As expected from the statistics of Callisto flares, and because the method used for detection is tailored towards sudden increases in the trigger rates, the majority of these flares are of type III, sometimes with several subsequent distinct emission spikes.

\begin{figure*}
\includegraphics[width=\textwidth]{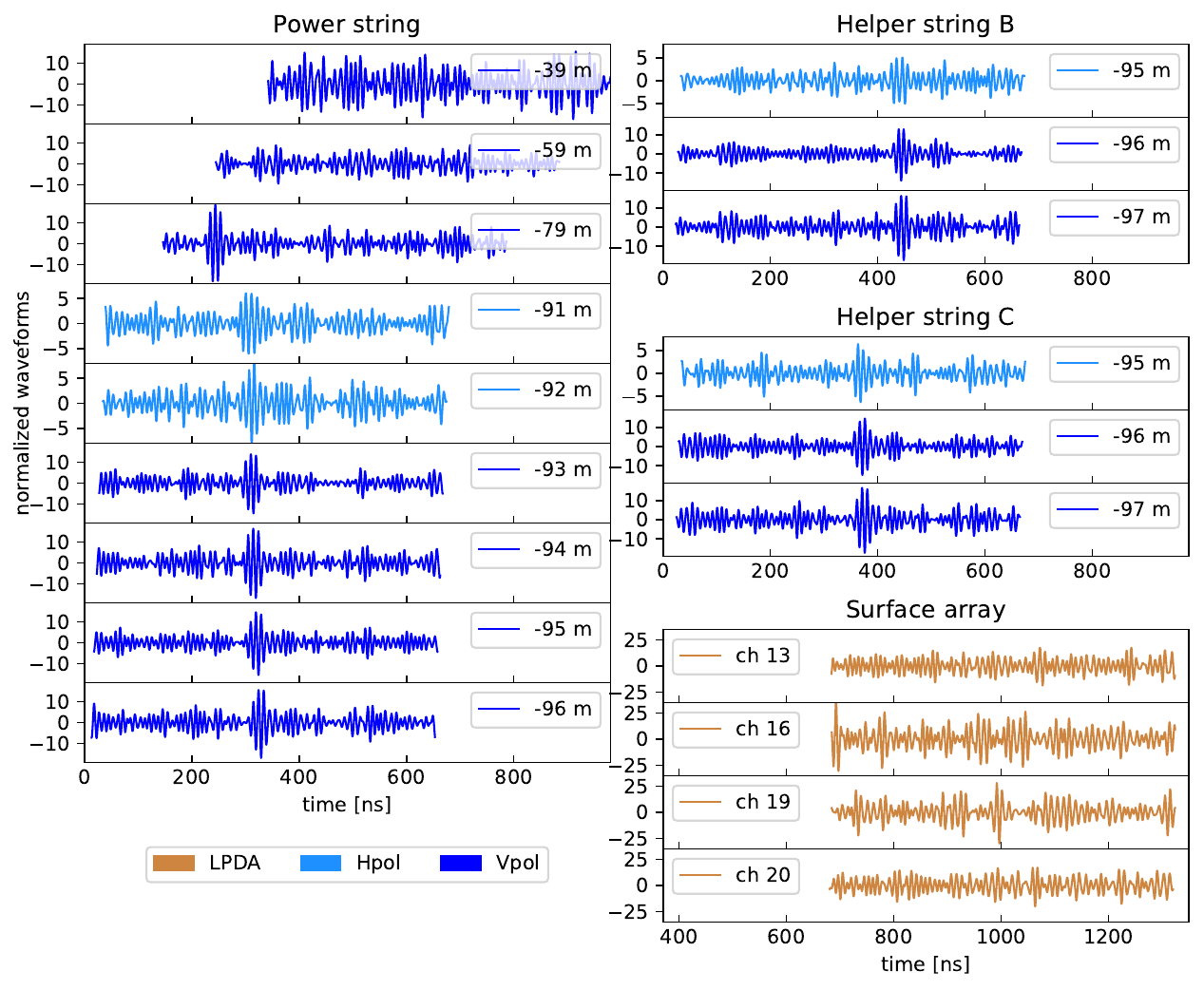}
\caption{Waveforms for one event recorded in \hbox{RNO-G} during the September 29, 2022 solar flare. The vertical axis shows the signal-to-noise ratio (SNR) in a given channel, normalized to the ambient thermal noise level pre-flare. Each channel's start time has been offset, relative to the Vpol at -96 m (power string), by an amount equal to the cable delay, to show the true signal arrival time at the location of that channel. The evident peak signal arrival time in the shallower Vpol at -79 m, for example, is approximately 80 ns earlier than Vpol at -93 m, as expected for a down-coming plane wave incident from the Sun, given the depths of the two channels. Signals in Vpols at -39 m and -59 m arrive too early in the buffer to be captured at the time when the deep channels initiate the DAQ trigger latch. Similarly, signals arrive at the surface LPDAs (orange) approximately one-half microsecond earlier than the deep in-ice channels; post-peak data are captured hundreds of nanoseconds after the actual signal has already swept through those channels. 
Surface \hbox{LPDA} channels 13, 16, and 19 point upwards, rotated by 120 degrees azimuthally (and therefore have different gains in the direction of the Sun), while \hbox{LPDA} channel 20 (shown for illustration) points downwards, and sees correspondingly less flare signal. For this reason, the remaining 5 downward-pointing \hbox{LPDA} channels have been omitted from this figure.}
\label{fig:EventWF}
\end{figure*}

\section{Reconstructing the flare direction}
\label{sec:reconstruction}
As opposed to data from solar spectrographs, such as those provided by HUMAIN, every vertical line in the dynamic spectrum from \hbox{RNO-G} is calculated from one waveform recorded in the time-domain, as shown in \autoref{fig:EventWF}. 
These same data therefore allow us to reconstruct the incident direction of the emission as observed with RNO-G. 
For the \hbox{RNO-G} neutrino mission, such a reconstruction is highly relevant to provide absolute pointing, calibration of the positions of the antennas, and an illustration of the achieved resolution on the signal arrival direction.

For the purpose of this section, we selected a subset of 28 `highly-impulsive' flares as identified by the SWAVES and Callisto correlation with \hbox{RNO-G} data, to study the reconstruction performance. These flares were chosen on the basis of the highest amplitude of the coherently summed waveform for the in-ice Vpol channels. 

\subsection{Expected position resolution and direction}

The expected solar reconstruction resolution is determined by three factors: knowledge about the location of the emission source on sky, the trajectory followed by the signals (``rays'') as they traverse the air-ice boundary and are refracted into the ice, and the \hbox{RNO-G} instrument characteristics. 

Typical solar flare spatial characteristics have been well-studied by dedicated radio-frequency observatories. Source locations on the Sun have been fixed by LOFAR, e.g., with typical resolutions of O(10'') in angle; such high precision is largely attributable to the large number ($\mathcal{O}(500)$) and spacing  ($\mathcal{O}(500)$ m) between receiver antennas used for interferometric reconstruction, affording strong constraints on the direction of incident signals. For RNO-G, the expected resolution is poorer, owing to (in order of importance): i) the meter-scale vertical, and 10-meter-scale horizontal separation between \hbox{RNO-G} receiver antennas, ii) the required correction for refraction at the air-ice boundary, iii) uncertainties in the positions of the antennas in-ice -- although the surface location of the holes is determined to within 5 cm using high-accuracy GPS, possible hole-tilting during drilling is not monitored, so lateral deviations of the in-ice antennas relative to the surface of order 10-20 cm are possible, iv) uncertainties in the refractive index profile in the ice. 

Since \hbox{RNO-G} has performed extensive measurements of all relevant cable delays (nominally quoted to a precision of several hundred picoseconds), uncertainties in propagation times through the DAQ are a sub-dominant effect. Similarly, the relative timing of wavefronts passing through two antennas is also determined to an accuracy of order 100 picoseconds; this is particularly true of solar flares, as the entire 500\,ns of a typical waveform trace contains an RF imprint of the emission. The time delay uncertainty results in an angular uncertainty of about 0.2 degrees, if antennas from different strings are used that are sufficiently distant from each other.

Since the solar flares saturate the trigger rate at several tens of Hz and, unlike neutrinos or air showers, the signals are not single sharp $\sim$1--10 ns duration peaks in the captured waveforms, it is unlikely that a single leading edge triggers all stations, which would allow for reconstruction across several RNO-G stations and provide an even better angular reconstruction. A search has revealed no temporal coincidences across all stations during the time of a solar burst and the two-fold coincidence matches the expectation for random coincidences. Therefore, we report only on single station reconstructions. 

The combination of these effects results in expected source location resolutions of order one degree, roughly corresponding to the solar-spot size in the sky at \hbox{RNO-G} frequencies (see Section \ref{sec:sun_rnog}). This value is comparable to previous South Polar solar flare reconstructions \citep{ARA:2018fkz}. Note that the hundreds-of-MHz frequency regime to which \hbox{RNO-G} is sensitive favors emission regions close to the surface; coronal plasma ejection away from the surface and into the photosphere favors longer-wavelength (decameter, e.g.) portions of the spectrum.

\begin{figure}
\includegraphics[width=\columnwidth]{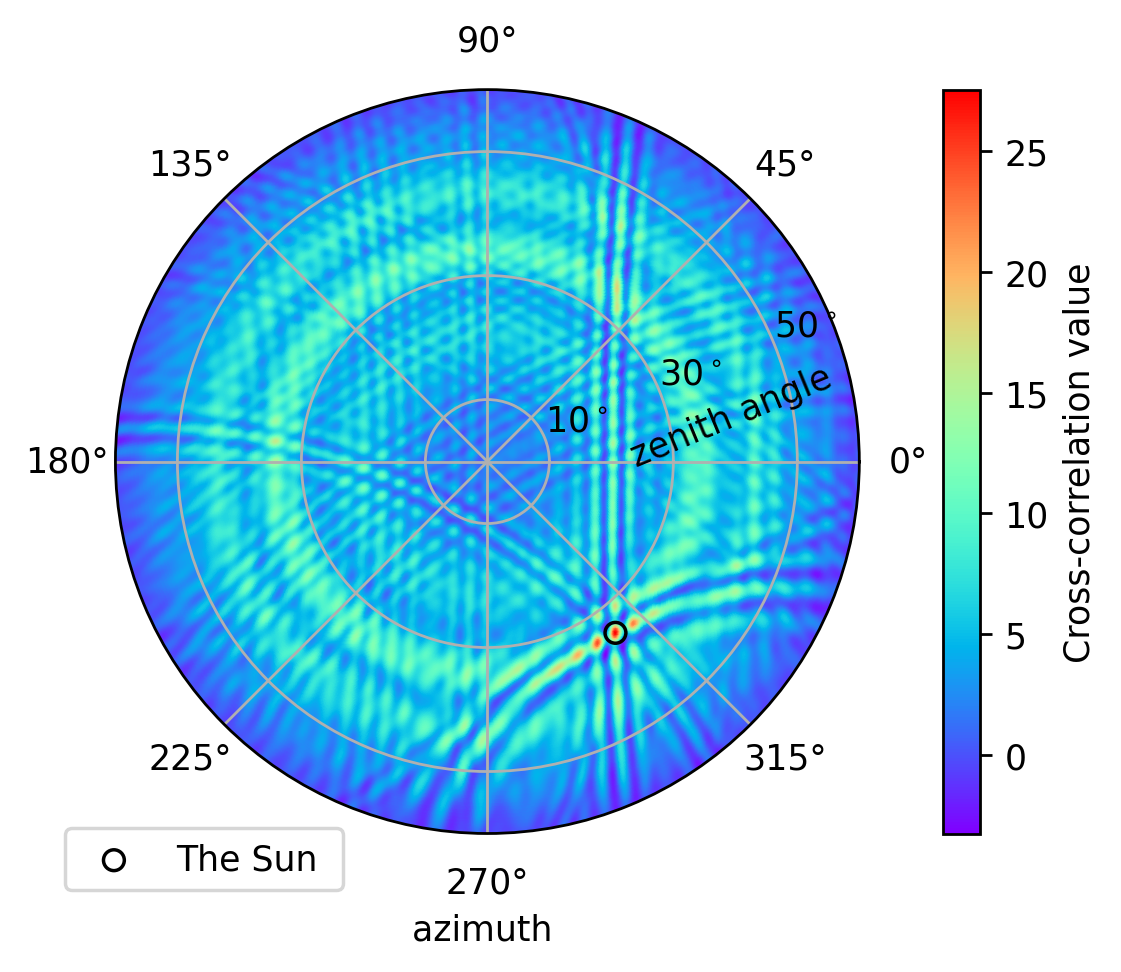}
\caption{Interferometric sky map for one set of station 11 waveforms recorded during the flaring Sun on September 29, 2022. Circle indicates the maximum pixel for this event, which is taken as the measured solar location on the sky (not corrected for refraction). Color scale indicates the cross-correlation value for a given pixel.}
\label{fig:IFM}
\end{figure}

\begin{figure}
\includegraphics[width=\columnwidth]{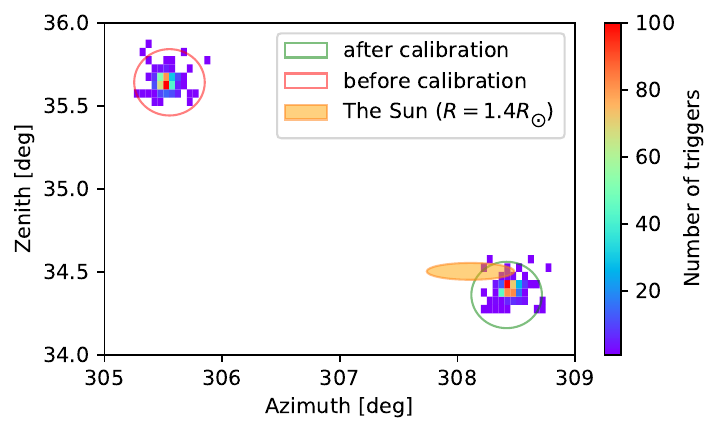}
\caption{Reconstructed source location for the solar flare of September 29, 2022, zoomed in as compared to \autoref{fig:IFM}. Reconstruction results both before and also after calibrating the antenna positions of station 23 are shown. The size of the Sun is chosen as the radio spot size of the Sun, corrected for refraction at the air-ice boundary.}
\label{fig:zenith_azimuth}
\end{figure}

\subsection{Interferometric reconstruction}
\autoref{fig:EventWF} shows the time-domain waveforms recorded during the solar flare on September 29th, 2022, which clearly illustrates the increasing delay of the voltage peak for the deeper antennas in a given string, and is the signature of a plane wave propagating downwards through the array.

It should be noted that, for the flares presented in this work, the firmware of the instrument was not capable to in-situ correct for the time-delays between \hbox{LPDA} and Vpol/Hpol channels due to the drastically different length of cables. This means that all antennas are read-out, but the signals are not causally related due to the travel time. Thus, no concurrent recordings in all channels are present for this year of observation and \hbox{LPDA} and Vpol/Hpol triggers have to be analyzed separately. 

The standard \hbox{RNO-G} source reconstruction codes, designed for both cosmic ray as well as neutrino reconstruction, are based on interferometry of the signals recorded in the time-domain. For this study, we use the software framework NuRadioReco \citep{Glaser:2019rxw} for the interferometric reconstruction of the signal arrival direction of solar flare events. The time delays that maximize the total cross-correlation between all possible Vpol (pairwise) combinations are directly translated into a most-likely signal arrival direction in both elevation and also azimuth. To calculate this total cross-correlation, we only sum over the pairwise cross-correlation between the Vpols at $\sim-80$\,m and below, since the expected arrival time for a downward propagating signal lies outside the recorded time-window for the shallower antennas.

A single interferometric map calculated for the waveforms depicted in \autoref{fig:EventWF} is shown in \autoref{fig:IFM}.
For each captured event, the brightest pixel in a zoomed ($\pm$5 degrees from the known solar location) interferometric map is recorded. The reconstructed source location, for all events recorded during the flare in one sample station, is presented in \autoref{fig:zenith_azimuth}. We observe a deviation, relative to the known solar position at a given time, of approximately 2 degrees in azimuth and one degree in elevation. The remaining \hbox{RNO-G} stations also show discrepancies of $\sim$1--2 degrees in both coordinates, which emphasizes the need for the calibration of the absolute pointing of RNO-G.

\subsection{Geometry calibration and systematic uncertainties}
Consistent with the interferometric reconstruction approach, antenna positions are calibrated in \hbox{RNO-G} relative to each other. Starting from nominal installation positions (known depth based on the measured length of the string, but less well-known azimuthal orientation due to absolute borehole position and inclination uncertainties), signals from embedded local calibration pulsers are used to initially refine the tabulated receiver antenna positions. Fine-tuning requires absolute pointing to distant external sources such as the Sun. 

Starting with the observed deviations between the reconstructed and the known solar positions, we have used a minimization code to extract the antenna coordinates which give the best match to the known solar location in the sky at any given time. For this minimization, we assume that the relative cable delays are known; the refractive index profile as function of depth $n(z)$ is drawn from \cite{Oeyen:2023eN}, although we later toggle the $n(z)$ profile by $\pm$1\%, and obtain consistent results. We also find that our results are relatively insensitive to cable delay uncertainties of order $\pm$0.2 ns. Our fitting procedure is significantly over-constrained -- for each of the 17 flares used for the minimization, there are approximately 50--500 reconstructable independent events, often spread across a time interval large enough that the Sun drifts 2--3 degrees in the sky.

This minimization procedure yields individual shifts of order 10--20 cm in (x,y) for the in-ice antennas, with much smaller shifts obtained in depth (z). Although the antennas were shifted individually in the minimization procedure, the fitted horizontal displacements of the antennas from one string were approximately the same, as expected. Sub-dividing the solar flares into two sub-samples significantly degrades the statistical precision, but allows us to estimate systematic errors that are approximately half the size of the calculated shifts (5 cm, obtained as the width of the distribution of the difference in post-calibration positions for the two subsamples). The obtained position shifts in (x,y) are somewhat larger than the expected uncertainty in the GPS-obtained antenna string positions, however, such shifts are commensurate with possible deviations from verticality during drilling and are in agreement with local pulser calibration results, as shown in  \citep{InstrumentPaper}. 

\begin{figure}
    \centering
    \includegraphics[width=\linewidth]{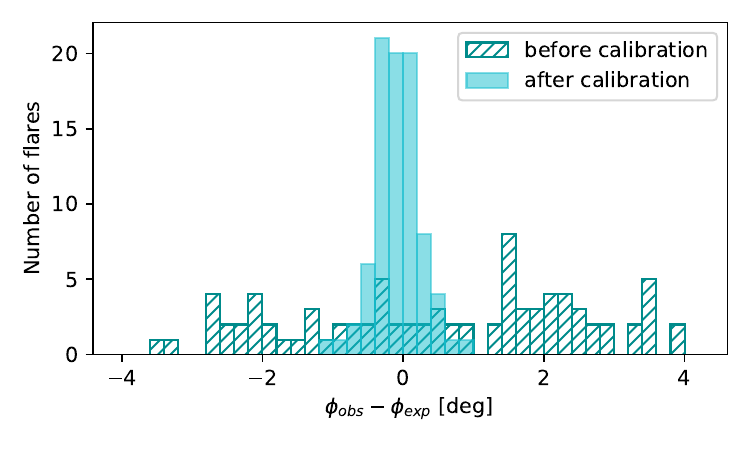}
    \includegraphics[width=\linewidth]{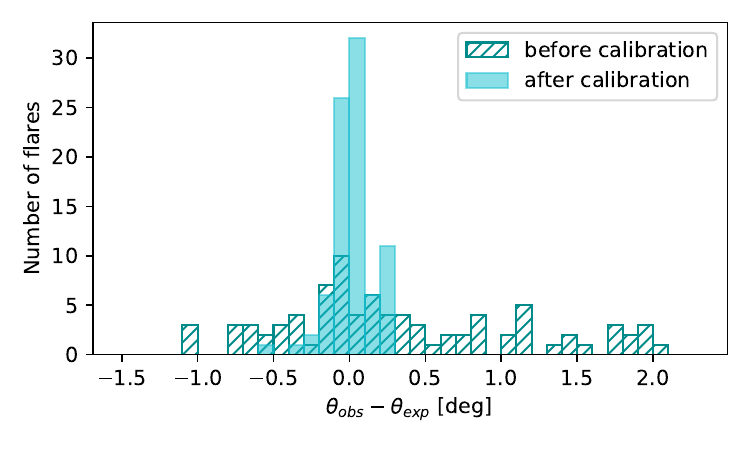}
    \caption{Reconstruction offset in azimuth $\phi_\mathrm{obs}$ (top) and zenith angle $\theta_\mathrm{obs}$ (bottom) observed after refraction by the deep antennas with respect to the arrival direction  expected from the centre of the Sun ($\phi_\mathrm{exp}$, $\theta_\mathrm{exp}$). The histogram contains 86 data points, one for each active station
    and 28 reconstructed highly impulsive flares (of which 58 events from 17 flares were used as calibration sample).}
    \label{fig:dphidtheta_per_flare}
\end{figure}

\begin{figure}
    \centering
    \includegraphics[width=\linewidth]{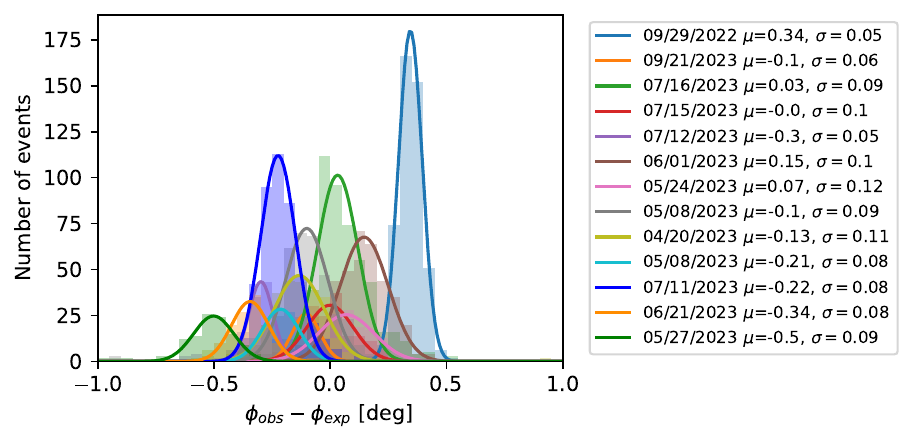}
    \includegraphics[width=\linewidth]{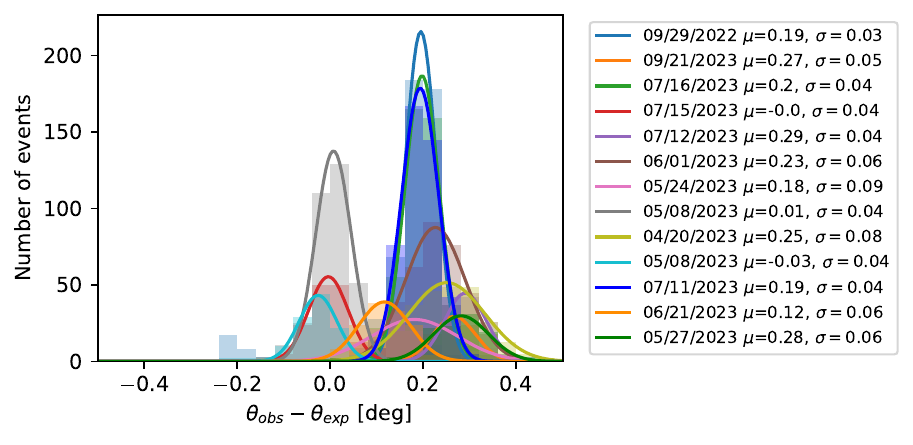}
    \caption{Reconstruction offset after calibration in azimuth $\phi_\mathrm{obs}$ (top) and zenith angle $\theta_\mathrm{obs}$ (bottom) observed after refraction by the deep antennas with respect to the arrival direction  expected from the centre of the Sun ($\phi_\mathrm{exp}$, $\theta_\mathrm{exp}$). Each flare of the calibration sample is indicated by a different color; offset and spread of the distributions are obtained via a Gaussian fit.}
    \label{fig:dphidtheta_post_calib}
\end{figure}

After calibration, both the solar azimuth and elevation reconstructions improve considerably relative to the nominal solar position. \autoref{fig:dphidtheta_per_flare} shows the distribution of the deviation in observed direction (after refraction into the ice) for all reconstructed flares and stations. The root-mean-square deviations after calibration for individual stations are 0.25$^\circ$ in azimuth $\phi$ and 0.12$^\circ$ in zenith angle $\theta$.
For one representative station, \autoref{fig:dphidtheta_post_calib} shows the distribution of reconstructed $\phi$ and $\theta$ for all individual events in our solar flare sample, after calibration. We observe that, for each station, the reconstructed precision in angular deviation is of order 0.1-0.2 degrees, albeit with comparable systematic offsets from zero. In \autoref{fig:dphidtheta_per_flare} and \autoref{fig:dphidtheta_post_calib} we have chosen to display $\phi$ and $\theta$ independently. The difference between $\phi$ and $\sin(\theta)\phi$, which would account for space angle, is smaller than a 25\% correction because the Sun is never close to the zenith at RNO-G. The overall reconstruction precision for solar flares approaches 15 arc-minutes.

\begin{figure}
    \includegraphics[width=0.49\textwidth]{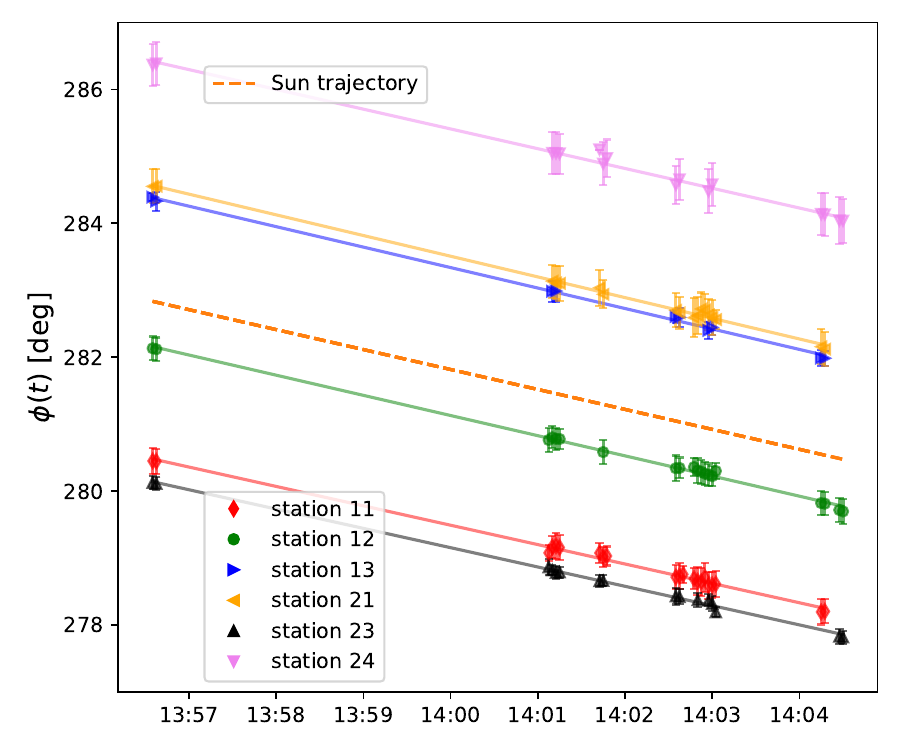}\newline
    \includegraphics[width=0.49\textwidth]{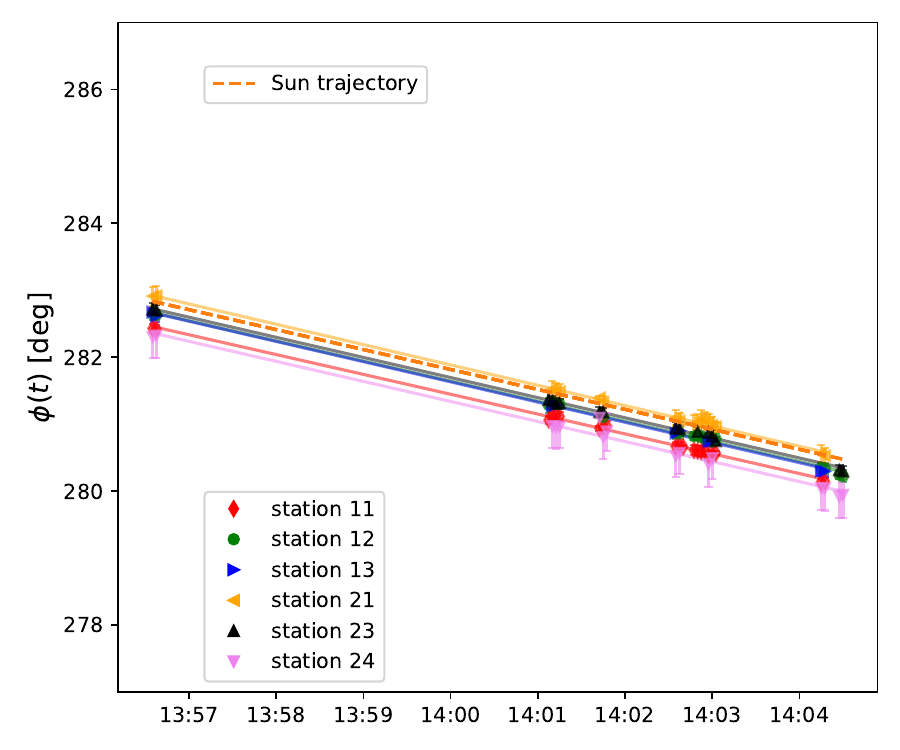}
    \caption{Reconstructed azimuth angle versus time for the solar burst recorded on July 12th, 2023 before (top) and after applying the calibration of the antenna positions (bottom). Markers indicate times at which the recorded signals allowed for an accurate pointing of the Sun. Also indicated is the expected trajectory of the Sun (dashed line, no markers). Due to a high trigger threshold, station 22 did not record enough data during the flare to reconstruct the source direction.}\label{fig:reco_simple_calibration2023_all}
\end{figure}

\begin{figure}
    \includegraphics[width=0.49\textwidth]{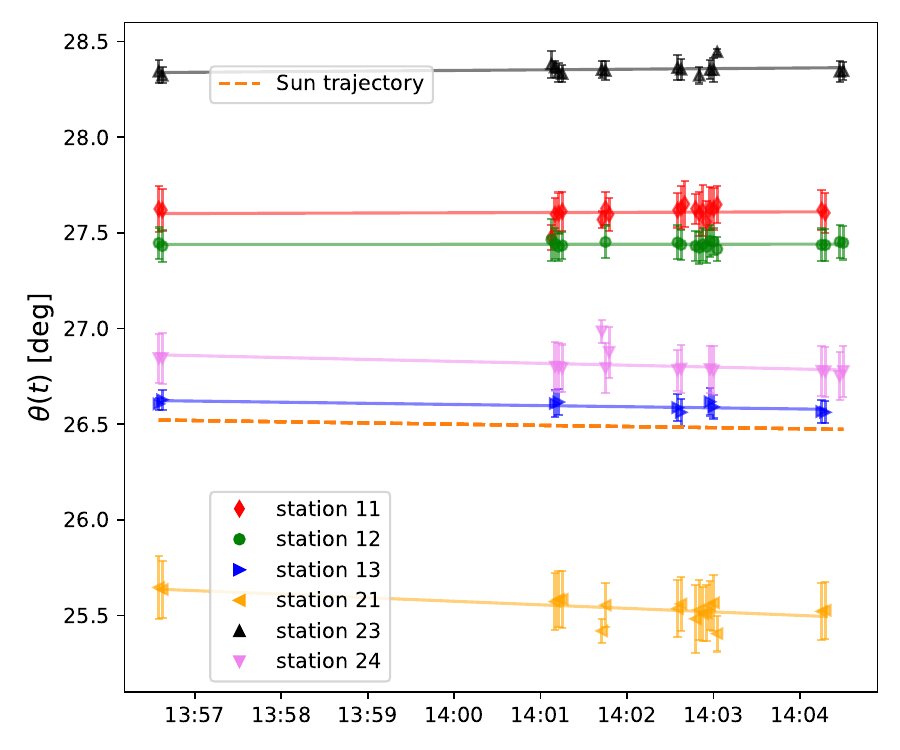}\newline
    \includegraphics[width=0.49\textwidth]{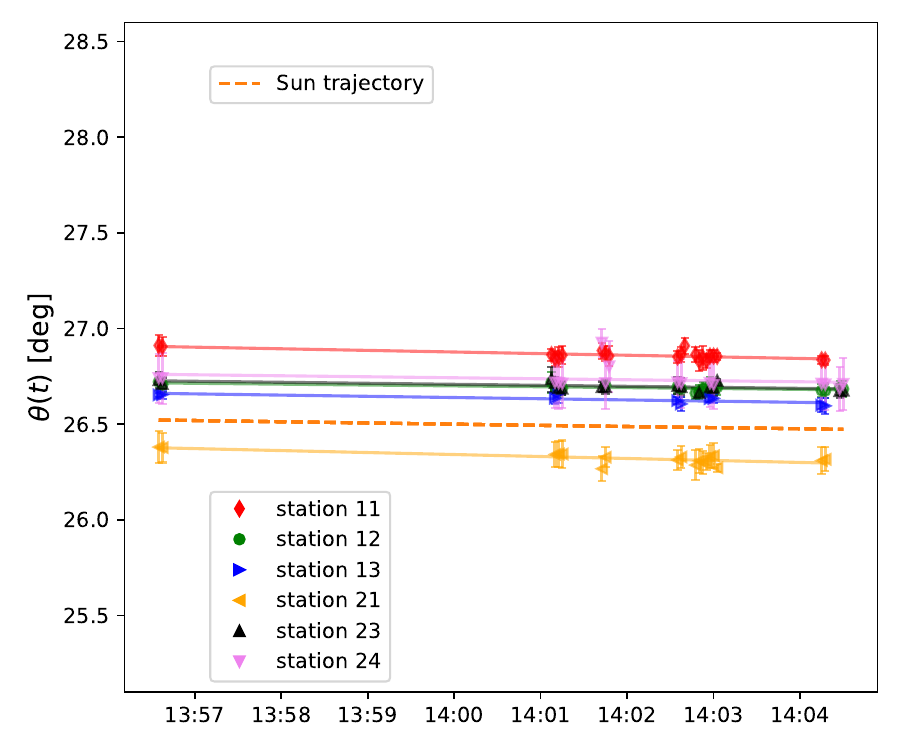}
    \caption{Same as \autoref{fig:reco_simple_calibration2023_all}, but for reconstructed zenith angle. Note that not all flares can be reconstructed in all stations.}
    \label{fig:reco_burst_vs_t}
\end{figure}

We also note that we can easily track the motion of the Sun in the sky, as shown over the 8-minute duration of the 12th of July, 2023 solar flare
(\autoref{fig:reco_simple_calibration2023_all}
and \ref{fig:reco_burst_vs_t}). As shown in \autoref{fig:zenith_azimuth} the Sun is comparable in size to the \hbox{RNO-G} angular resolution. Although solar flares typically do not originate from the center of the Sun (resulting in a corresponding inherent uncertainty in the source location), we assume a central emission point in the absence of more precise information. After further refinement of the calibration and with more input from other observatories, we anticipate that a true offset from the solar center (as visible in \autoref{fig:zenith_azimuth}) may be measured for future flares. We note that, although each individual fitted event is constrained to reconstruct at the location of the Sun by fiat, the calibrated antenna coordinates will vary flare-to-flare, event-to-event and station-to-station. For the final comparison of the known solar location with our reconstruction, we take the average positional shift per channel, which, when applied to our flare sample, will no longer necessarily reconstruct to the center of the Sun.

\section{Time-domain characteristics of signal shapes recorded during solar flares}

\begin{figure}
    \includegraphics[width=\columnwidth]{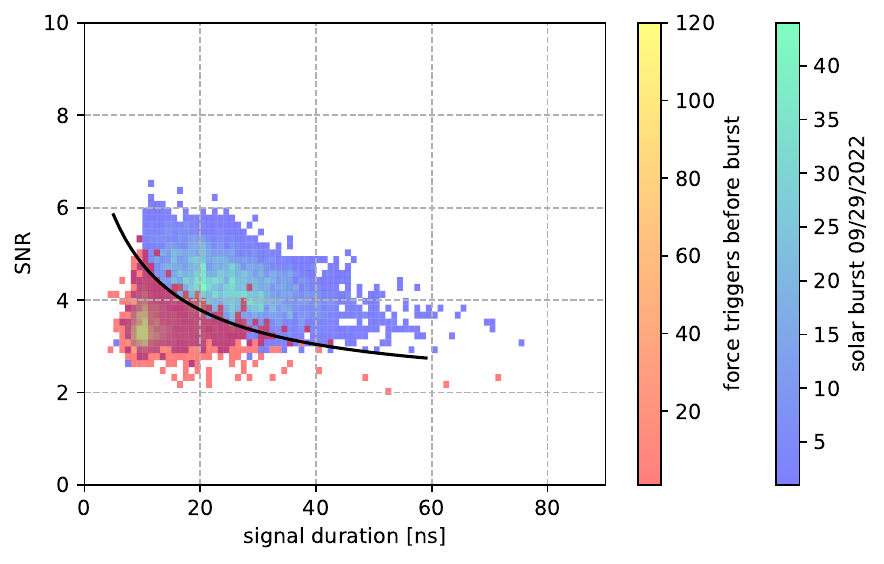}
    \includegraphics[width=\columnwidth]{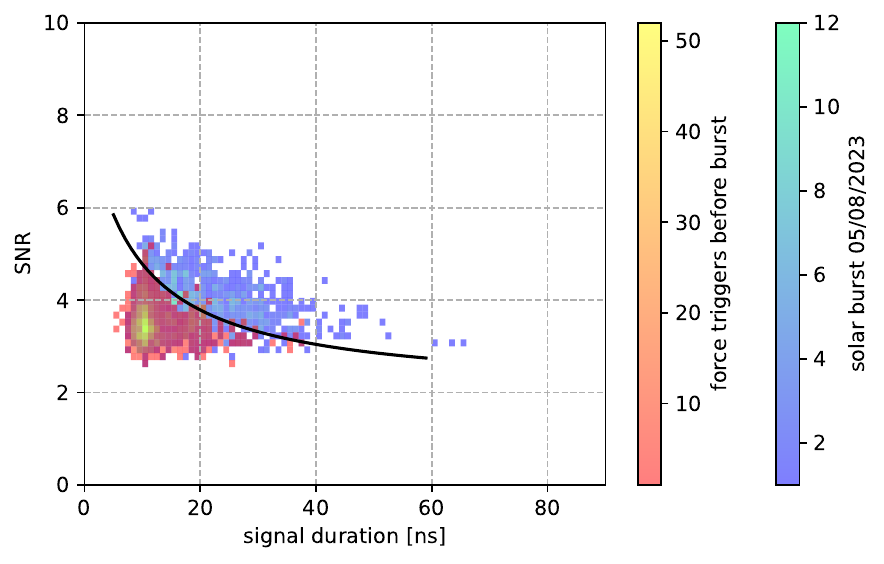}
    \caption{Illustration of the impulsivity of the signals recorded. Signal-to-noise ratio (SNR) is shown as function of signal duration. The figure compares the noise signal recorded during normal operations (periodic forced triggers) with RF triggered signals recorded during the solar burst of September 29th, 2022 (top) and May 8th, 2023 (bottom). The area above the black line admits less than 1\% of forced triggers, and is used to estimate the number of solar flare signals that are dissimilar from noise. }
    \label{fig:s_shape}
\end{figure}

As evident in \autoref{fig:EventWF}, we observe transient (of order 10 ns) signals in our waveforms, indicative of acceleration at the source, on a similar time scale. We stress, however, that the recorded time-domain waveforms include dispersive effects from the ionosphere (discussed below), the antenna itself, and the data acquisition system, for which we have not corrected here. To quantify how often such impulsive signals are apparent in our solar flare candidate waveforms, we constructed a two-dimensional plot of peak signal-to-noise ratio in an observed waveform vs.\ the signal duration, defined as the time duration between the signal onset (in this case, the time at which the signal voltage envelope exceeded half-maximum) and the signal falling edge (defined as the time at which the recorded signal voltage fell below half-maximum). The signal-to-noise ratio is calculated as the ratio of the signal amplitude to the voltage standard deviation of the waveform containing the signal. \autoref{fig:s_shape} shows this two-dimensional plot, and also shows the cut (black line) corresponding to those events which are more readily interferometrically reconstructed (blue) vs.\ those events for which the peak in the interferometric map is considerably weaker. Continuous-wave background generally exhibits a signal-to-noise ratio of less than 2 and therefore falls below the cut line. Although our triggered data sample in the absence of solar flares is dominated by events close to the thermal noise floor, that sample also include impulsive anthropogenic backgrounds and therefore are less useful, compared to forced triggers, which are used as a reference thermal noise sample. With this criterion, approximately 1/3 of the event triggers recorded during known solar flares produce observable impulsive signals in our receivers. 

We have considered the science that drives the observed nanosecond-scale structures; to our knowledge there are no previous solar physics publications using solar flares recorded from Earth with this timing resolution. Given the absence of data, there has been very little  modeling of extremely short time-scale effects. 
Typical `fast' solar studies e.g.~\cite{2003A&A...407.1115M}. \cite{Dabrowkski2015} have revealed microsecond-scale structures in a particularly rapid type of narrowband solar burst (\emph{dm-spikes}) and a drifting pulsation structure of the same time-scale, but at much higher frequencies of 0.8 to 2.0 GHz. Is is therefore difficult to directly translate the implications of their findings (that these are the radio signature of a cascade of interacting plasmoids or more generally due to details in the plasma physics of the Sun) to RNO-G. By releasing RNO-G solar data, we encourage theoretical work to consider the extremely short time-scales. 

In principle, it should be possible to extract polarization from the time-domain signals, as has been demonstrated for cosmic-ray signals and in-ice sources \citep{ARIANNA:2020zrg,Arianna:2021lnr}. However, for broadband signals like those observed in RNO-G, ionospheric dispersion (as function of frequency) may play an important role for signals originating from outside the Earth's atmosphere \citep{1983A&A...120..313S}. In particular, at the low frequencies at which \hbox{RNO-G} operates, dispersive delays across the band can add up to several tens to hundreds of nanoseconds, depending on the state of the ionosphere. Without additional information about the ionosphere, it is therefore challenging to de-disperse the pulse and also correct for any polarization rotation that may be introduced during ionospheric propagation. 
For time integrated quantities the polarization calculation should be possible~\citep{Lofar_solar_flares} (as dispersion will mostly cancel out), however, this does not reveal characteristics of individual pulses, as with cosmic ray or neutrino signals. 
Considerable additional work will therefore be needed to assess the utility of the \hbox{RNO-G} data for polarization measurements of solar emission, exploiting the entire bandwidth for the best results.

\section{Conclusions}
RNO-G radio observations of solar flares are both interesting, in that they reveal nanosecond-scale broadband acceleration mechanisms, and also useful, in that they afford a powerful, infinite-distance calibration source at a well-defined location on the sky. From a calibration standpoint, as demonstrated herein, after optimizing the coordinates of the receiver antennas, using the known position of the Sun in the sky as a constraint, both azimuthal and elevation reconstruction precision for above-surface sources approaches 15 arc-minutes. We have also shown in this paper that using the Sun as known position, we can achieve sub-degree absolute pointing on the signal arrival direction for all \hbox{RNO-G} stations viewing above-ice sources. Both reconstruction precision and absolute pointing are clearly adequate for cosmic ray reconstruction as well as neutrino reconstruction, although verification of the latter awaits reconstruction of dedicated in-ice transmitter sources, at ranges of hundreds of meters. The directional information of charged cosmic rays is intrinsically already smeared by magnetic fields to more than a degree from their sources and for neutrinos current arrival direction reconstructions estimate only a small fraction (less than 10\%) to be reconstructed better than $1^{\circ}$ statistical precision due to the dominant additional uncertainties from determining the position on the Cherenkov cone, not yet including systematic uncertainties. 

We note that, in the approximately 50 station-months of \hbox{RNO-G} livetime, there have been a total of 75 solar flares identified in the \hbox{RNO-G} data stream. These are readily characterized by a sharp enhancement in both station trigger rates and also station signal power. These characteristics make it unlikely that solar flares are a considerable background to neutrino searches with RNO-G, which illuminate only one or two \hbox{RNO-G} stations with single, non-repeating pulses. 

In contrast to the large \hbox{RNO-G} solar flare data sample already accumulated, the Radio Ice Cherenkov Experiment (RICE, \cite{RICE:2001ayk}), which ran continuously at the South Pole from 2000-2011 (comprising 144 station-months of livetime), observed only the (extremely bright) X-28 November 4, 2003 flare. Although \hbox{RNO-G} fortunately has been commissioned close to solar maximum, we attribute the higher efficiency of \hbox{RNO-G} to the more favorable solar viewing angle -- the more inclined signal angles typical of the South Pole result in more reflected signal power at the surface, and therefore suppressed efficiency of the in-ice triggers typical of experiments such as RICE. 

RNO-G is currently still under construction and we have reported here only data from the first year of 7 stations. \hbox{RNO-G} is scheduled to reach 35 stations with a total of 840 antennas by 2026. Adapting reconstructions across multiple stations and carefully adjusting the trigger rate  will lead to extremely high resolution spectra, as well as precision location of solar flares almost as a by-product of the ongoing neutrino observation program. 
Future modifications to the trigger contemplate real-time suppression of trigger-threshold adaptation to allow a non-constant trigger rate for candidate solar flares, which would then instead maintain a constant trigger signal threshold and better articulate the tail end of the solar light curve, and provide considerably larger data sets, as the Sun approaches solar maximum in the summer of 2024.

\section*{Acknowledgements}
We are thankful to the support staff at Summit Station for making \hbox{RNO-G} possible. We also acknowledge our colleagues from the British Antarctic Survey for building and operating the BigRAID drill for our project. We would like to thank Christian Vocks for useful discussions regarding solar physics.

We would like to acknowledge our home institutions and funding agencies for supporting the RNO-G work; in particular the Belgian Funds for Scientific Research (FRS-FNRS and FWO) and the FWO programme for International Research Infrastructure (IRI), the National Science Foundation (NSF Award IDs 2118315, 2112352, 2111232, 2112352, 2111410, and the collaborative awards 2310122 through 2310129), and the IceCube EPSCoR Initiative (Award ID 2019597), the German research foundation (DFG, Grant NE 2031/2-1), the Helmholtz Association (Initiative and Networking Fund, W2/W3 Program), the Swedish Research Council (VR, Grant 2021-05449 and 2021-00158), the Carl Trygers foundation (Grant CTS 21:1367), the University of Chicago Research Computing Center, and the European Union under the European Unions Horizon 2020 research and innovation programme (grant agreement No 805486), as well as (ERC, Pro-RNO-G No 101115122 and NuRadioOpt No 101116890). 
We thank Fachhochschule Nordwestschweiz (FHNW), Institute for Data Science in Brugg/Windisch, Switzerland, for hosting the e-Callisto network. 

\bibliographystyle{cas-model2-names}
\bibliography{bibliography}

\end{document}